\documentclass[pdflatex,referee,sn-nature]{sn-jnl}

\usepackage[justification=justified, font=small]{caption}
\usepackage{graphicx}%
\usepackage{multirow}%
\usepackage{amsmath,amssymb,amsfonts}%
\usepackage{amsthm}%
\usepackage{mathrsfs}%
\usepackage[title]{appendix}%
\usepackage{xcolor}%
\usepackage{textcomp}%
\usepackage{manyfoot}%
\usepackage{booktabs}%
\usepackage{algorithm}%
\usepackage{algorithmicx}%
\usepackage{algpseudocode}%
\usepackage{listings}%
\usepackage{mathtools}
\usepackage[resetlabels]{multibib}
\usepackage{pdfpages}
\usepackage{array}
\usepackage{xr}
\usepackage{xr-hyper} 
\usepackage{ulem}

\usepackage{hyperref} 

\hypersetup{
    colorlinks=true,
    linkcolor=blue,
    filecolor=blue,      
    urlcolor=blue,
    citecolor=blue,
    pdftitle={maintext},
    }

\newcommand{\rev}[1]{{\color{black} #1}}
\newcommand{\abs}[1]{{\lvert #1 \rvert}}
\newcommand{\reb}[1]{{\color{blue} #1}}

\newcommand{\meths}[0]{\hyperref[meth]{Methods}}

\makeatletter
\newcommand*{\addFileDependency}[1]{
  \typeout{(#1)}
  \@addtofilelist{#1}
  \IfFileExists{#1}{}{\typeout{No file #1.}}
}
\makeatother

\newcommand*{\myexternaldocument}[1]{%
    \externaldocument{#1}%
    \addFileDependency{#1.tex}%
    \addFileDependency{#1.aux}%
}

\myexternaldocument{arxiv_CoherentDetection_SI_Arxiv_v2}

\title{\textbf{Graphene-enabled coherent \rev{\rev{(sub-)}}terahertz wave detection and thin film thickness determination}}

\date{\today}

\author[1]{\fnm{Ronny} \sur{de la Bastida}}

\author[1]{\fnm{Enzo} \sur{Rongione}}
\equalcont{These authors contributed equally to this work.}

\author[2]{\fnm{Karuppasamy Pandian} \sur{Soundarapandian}}
\equalcont{These authors contributed equally to this work.}

\author[3]{\fnm{Ioannis} \sur{Vangelidis}}

\author[1]{\fnm{Anand} \sur{Nivedan}}

\author[1]{\fnm{David} \sur{Saleta Reig}}

\author[4]{\fnm{Kenji} \sur{Watanabe}}

\author[5]{\fnm{Takashi} \sur{Taniguchi}}

\author[3,6]{\fnm{Elefterios} \sur{Lidorikis}}

\author[2,7]{\fnm{Frank~H.~L.} \sur{Koppens}}

\author[2]{\fnm{Sebastián} \sur{Castilla}}

\author*[1,8]{\fnm{Klaas-Jan} \sur{Tielrooij}}\email{k.j.tielrooij@tue.nl}

\affil[1]{\orgname{Catalan Institute of Nanoscience and Nanotechnology (ICN2), CSIC and BIST}, \orgaddress{\street{Campus UAB, Bellaterra}, \postcode{08193}, \state{Barcelona}, \country{Spain}}}

\affil[2]{\orgname{ICFO - Institut de Ciències Fotòniques, The Barcelona Institute of Science and Technology}, \postcode{08860}, \state{Castelldefels (Barcelona)}, \country{Spain}}

\affil[3]{\orgdiv{Department of Materials Science and Engineering}, \orgname{University of Ioannina}, \postcode{45110}, \city{Ioannina}, \country{Greece}}

\affil[4]{\orgdiv{Research Center for Electronic and Optical Materials}, \orgname{National Institute for Materials Science}, \orgaddress{\street{1-1 Namiki}, \city{Tsukuba}, \postcode{305-0044}, \country{Japan}}}

\affil[5]{Research Center for Materials Nanoarchitectonics, National Institute for Materials Science, 1-1 Namiki, Tsukuba, 305-0044, Japan}

\affil[6]{University Research Center of Ioannina (URCI), Institute of Materials Science and Computing, 45110, Ioannina, Greece}

\affil[7]{\orgname{ICREA - Institució Catalana de Recerca i Estudis Avançats}, \postcode{08010}, \state{Barcelona}, \country{Spain}}

\affil[8]{\orgdiv{Department of Applied Physics}, \orgname{TU Eindhoven}, \orgaddress{\street{Den Dolech 2}, \city{Eindhoven}, \postcode{5612 AZ}, \country{Netherlands}}}

\abstract{\textbf{Phase-sensitive terahertz (THz) detection enables applications ranging from astronomy to non-destructive testing. However, current THz detectors lack phase sensitivity, unless they are combined with external interferometers, or through photomixing. This implies a large footprint and sensitive dependence on alignment.  Here, we demonstrate a graphene-enabled, on-chip, integrated \rev{(sub-)}THz detector-interferometer with optical cavity and antenna, exhibiting high sensitivity to the phase of incident THz light. We exploit this by determining the thickness of thin films placed in front of the detector-interferometer, obtaining a deep sub-wavelength thickness accuracy of a few micrometer, while we predict that an \rev{improved} accuracy is within reach. 
This is relevant for a range of industrial application domains, including automotive, construction, and health. 
We furthermore achieve a record-high external responsivity – \rev{considering bias-free graphene-based \rev{\rev{(sub-)}}THz detectors} – of \rev{172} mA/W and a noise-equivalent power of \rev{26} pW$~\rm{Hz}^{-1/2}$. This performance is due to enhanced absorption at the resonant cavity mode around 89 GHz, in agreement with multi-physics simulations. These results pave the way to exploiting coherent wave detection in the \rev{\rev{\rev{(sub-)}}}THz regime with utility in  spectroscopy, next-generation wireless communication, and beyond.
}}

\keywords{terahertz, photodetection, graphene, interferometer, thickness determination}

\begin{document}
\maketitle

\section*{Introduction}

Coherent wave detection, which exploits phase information instead of the amplitude or intensity of incident light, is an essential tool that lies at the core of many scientific and engineering fields. A common way to implement coherent wave detection is through wave interference~\cite{taimre_laser_2015,rakic_sensing_2019}. Phase-sensitive detectors and interferometers in the THz regime are particularly desirable for a wide range of applications, where the use of more conventional electromagnetic spectrum ranges is limited~\cite{tonouchi_cutting-edge_2007,samizadeh_nikoo_electronic_2023}. For example, in astronomy, the detection of specific cosmic THz radiation allows diving into the formation of planets and galaxies in the far regions of the Universe~\cite{coldUniverse,DetSpace,Lara-Avila2019}. Coherent THz detection also reinforces security applications~\cite{appleby_millimeter-wave_2007,Valzania:19,li_high-throughput_2023} and non-destructive testing in the automotive, building and construction, and electronics industries. A prime example is the determination of the thicknesses of thin, visible-opaque coatings and films, such as car paint layers~\cite{Pfeiffer:18,park_review_2019,Ellrich2020,liebermeister_terahertz_2021}. Coherent THz detection can also lead to a speed-up of wireless telecommunication systems~\cite{nagatsuma_advances_2016,Marconi2021}, where phase-sensitive data transfer schemes are combined with a high-frequency carrier wave~\cite{kakande_multilevel_2011}, and in quantum key distribution systems~\cite{QUAM,Wang2021}.
\\

In the most common coherent THz detection approaches, the amplitude and phase of weak THz signals are measured through heterodyne and homodyne detection~\cite{seeds_coherent_2013,harter_siliconplasmonic_2018,Lara-Avila2019,Thomson2024}. This is the case for THz time-domain spectroscopy, where THz pulses are detected through electro-optic sampling in non-linear crystals followed by balanced photodetection~\cite{valdmanis_picosecond_1982,wu_ultrafast_1996,jiang_electro-optic_1998}. This is also the case for continuous wave THz spectroscopy using photomixers~\cite{Liebermeister2021,ben-atar_amplitude_2025}. Coherent THz detection through interferometry typically relies on external optical setups, such as Michelson and Fabry-P\'{e}rot interferometers, after which a photodetector registers the interference pattern of the mixed waves~\cite{Rakic:13,NGUYEN201419}. These approaches have a large footprint, significant power consumption, and a rather high cost, while being sensitive to (mis)alignment. A chip-sized, heterodyne-free THz detector with phase sensitivity would overcome these issues. 
\\

\begin{figure*}[!htp]
    \centering
    \includegraphics[width=\linewidth]{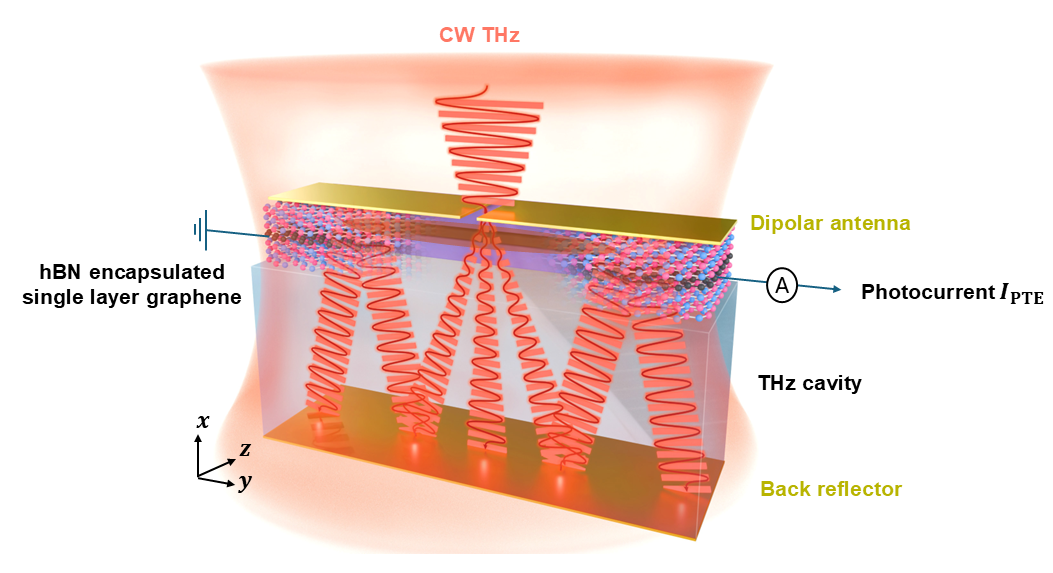}
    \caption{\textbf{Schematic of the integrated graphene-based \rev{(sub-)}THz detector-interferometer.} The device contains hBN-encapsulated graphene underneath a dipolar antenna. The photo-active region is the graphene channel below the gap of the dipolar antenna. This channel has side contacts (not shown) that enable us to measure photocurrent. The antenna and the metallic back mirror form an optical cavity, where incident (THz) light can travel back and forth and interfere. The \rev{(sub-)}THz cavity consists mostly of silicon.}
   \label{fig1}
\end{figure*}

Here, we introduce an on-chip integrated graphene \rev{\rev{(sub-)}}THz detector that is simultaneously an interferometer through its combination with a vertical \rev{\rev{(sub-)}}THz cavity that is formed between a metallic antenna and a metallic back mirror (see Fig.~\ref{fig1}). We demonstrate that the absorption enhancement in the cavity, together with the strong photo-thermoelectric response of graphene, gives rise to an external responsivity of \rev{172} mA/W and an external noise-equivalent power of \rev{26} pW$~\rm{Hz}^{-1/2}$. The detector does not require any bias voltage, resulting in minimal power consumption. Because of the large responsivity, the interference effects in the \rev{internal} cavity, \rev{the external emitter-detector cavity,} and the ultrathin photoactive layer formed by the graphene, this integrated \rev{\rev{(sub-)}}THz detector-interferometer enables highly phase-sensitive \rev{\rev{(sub-)}}THz photodetection\rev{, with a phase sensitivity below 0.02 rad}. We demonstrate the utility of this phase sensitivity by determining the thickness of optically opaque materials, in particular silicon, paper, kapton, and polypropylene. The thickness accuracy we achieve is a few micrometers, which is three orders of magnitude smaller than the wavelength. This accuracy can be improved to the nanometre regime by increasing the THz power and phase stability of the source. Beyond thickness determination, this coherent THz detector-interferometer opens up a myriad of possibilities for technologically relevant applications, for example wireless communications relying on phase-encoded protocols. 
\\

\section*{Results}

\subsection*{Device and signatures of interference}\label{sec2} 

We fabricated \rev{four} graphene-based detector\rev{s} that include a built-in vertical optical cavity, which leads to multiple reflections and constructive and destructive wave interference. The choice for graphene is based on its large Drude absorption~\cite{koppens_photodetectors_2014,THz_2D_rev} and the promising THz photodetection performance that has recently been demonstrated~\cite{vicarelli_graphene_2012,ShurBolometric, Bandurin2018, viti_hbn-encapsulated_2020,DNotorio2024,yao_configurable_2024,soundarapandian_high-speed_2024}. The atomically thin photoactive region furthermore minimizes the broadening of interference peaks. We optimized the detector specifically for the photo-thermoelectric effect, which simultaneously offers high sensitivity, fast response, broad spectral response, and passive operation~\cite{castilla_fast_2019,muench_waveguide-integrated_2019,schuler_high-responsivity_2021,asgari_chip-scalable_2021,koepfli_controlling_2024,castilla_electrical_2024,ludwig_terahertz_2024}. The detector\rev{s} contain a dipolar antenna that funnels \rev{(sub-)}THz radiation to the photoactive graphene area, and its two branches serve as electrostatic gates to create a junction in the graphene channel~\cite{Gabo_Jarillo_HotCarrier,castilla_fast_2019}. The antenna also serves as one of the two reflecting surfaces of the optical cavity, while the other surface consists of a metallic back mirror. In between the two reflecting surfaces, there is mainly silicon with a thickness of 279 µm, in addition to hBN-encapsulated graphene, which is directly below the antenna, as shown in Fig.~\ref{fig1} (see~\meths~and Supp.~Fig.~\ref{S_fab} for details). \rev{All the results in the main text belong to Device 1, unless explicitly stated. Supplementary Figs. S12-13 show the results obtained from Devices 2-4.}
\\

\begin{figure*}[!htp]
    \centering
    \includegraphics[width=\linewidth]{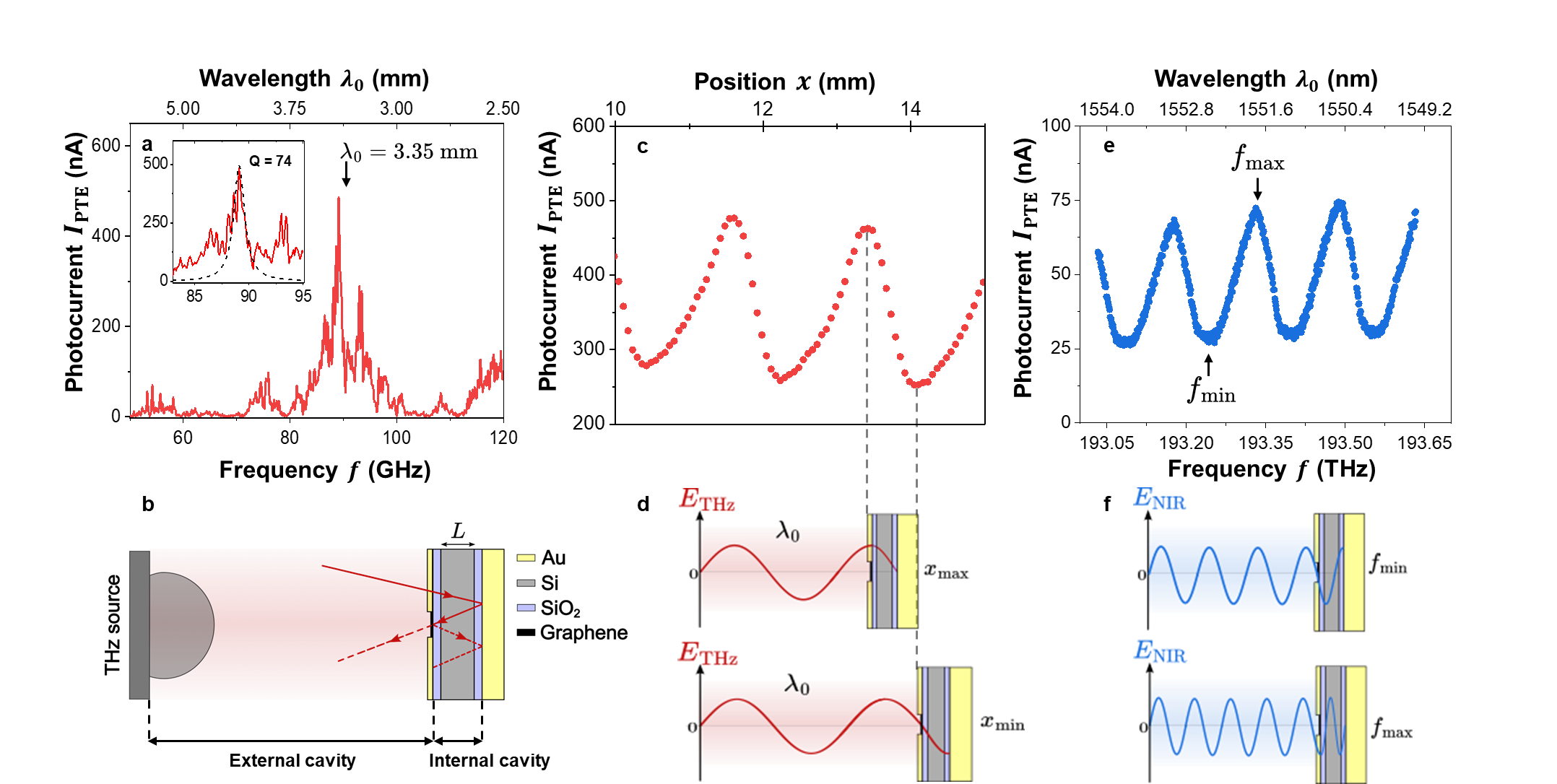}
    \caption{\textbf{Spectral and spatial signatures of interference.} \textbf{a,} Measured photocurrent as a function of frequency $f$ (wavelength $\lambda_0$), where a main peak occurs at 89 GHz (3.35 mm). The inset presents a zoom of the main peak with a Lorentzian fit (dotted black line). \textbf{b,} Schematic of the relevant internal and external cavities. The photocurrent peak corresponds to the situation where reflections in the internal cavity lead to constructive interference at the graphene location. The thinnest cavity for which this occurs is when $L = \frac{\lambda_{\rm 0}}{4 n_{\rm THz}}$. \textbf{c,} Measured photocurrent at 89 GHz while moving the detector along the light propagation direction $x$. The oscillatory pattern has a maximum (minimum) at position $x_\text{max}$ ($x_\text{min}$), with $\abs{x_\text{max} - x_\text{min}} = \lambda_0/4$. \textbf{d,} Schematic of the experimental geometry when moving the detector in the $x$-direction. Maxima (minima) occur when the incident field at the graphene position has a maximum (minimum). \textbf{e,} Measured photocurrent as a function of frequency in the near-infrared (NIR) range, showing periodic oscillations with maxima $f_\text{max}$ and minima $f_\text{min}$. \textbf{f,} Representation of the interference conditions at telecom frequencies, where the frequency spacing is determined by the internal cavity via $\abs{f_\text{max} - f_\text{min}} = c/4n_{\rm NIR} L$. }
    \label{fig2}
\end{figure*}

We first characterize our device by measuring the photocurrent $I_\text{PTE}$ as a function of frequency $f$ using a commercial continuous-wave THz setup (see~\meths). The results in Fig.~\ref{fig2}\reb{a} show a strong photocurrent peak at $f_{\rm peak} =$ 89 GHz, corresponding to a free-space wavelength of $\lambda_0$ = 3.35 mm. We ascribe this to positive interference at the location of the photoactive graphene area, which occurs for a cavity with a length $L$ equal to $\frac{\lambda_0}{4 n_{\rm THz}}$ (see Supp.~Fig.~\ref{S_cavity} for details). This gives a THz refractive index of $n_{\rm THz} =$ 3.1, which is close to the expected refractive index of silicon of around 3.5 in the THz range~\cite{franta_temperature-dependent_2017}. Examining the width of the photocurrent peak $f_{\rm width}$, we extract a quality factor of $Q=f_\text{peak}/f_\text{width}$ = 74 for the internal cavity formed between the metallic antenna and the metallic back mirror. In addition to the main photocurrent peak, we observe periodic peaks around the central frequency. We attribute these to the external cavity that is formed between the detector and the THz source, as shown in Fig.~\ref{fig2}\reb{b}. We will discuss this in more detail in the subsection that describes the finite-difference time-domain simulations.
\\

To further study interference effects in our device, we map the occurrence of constructive and destructive interference by moving the detector position along the THz propagation axis (out-of-plane direction $x$, see Fig.~\ref{fig1}) for incident light at 89 GHz, see Fig.~\ref{fig2}\reb{c}. The result looks like a typical interferogram with maxima and minima. The periodicity of the signal is 1.79 mm (see Supp.~Fig.~\ref{S_simus}), which is very close to $\lambda_0/2$. For incident light with a frequency of 160 GHz, we observe a similar interference pattern, now with a periodicity of 0.98 mm, which is again very close to $\lambda_0/2$ (see Supp.~Fig.~\ref{S_simus}). We therefore ascribe the observation of the occurrence of maxima and minima in photocurrent to the occurrence of maxima and minima in the intensity of the electric field of the incident \rev{(sub-)}THz light at the graphene position, as shown in Fig.~\ref{fig2}\reb{d}. This confirms that our detector works as an interferometer. 
\\

As final evidence of interference, we measure the photocurrent response using a tunable telecom laser in the wavelength range between 1549 and 1554 nm (see Supp.~Fig.~\ref{S_telecom}). Figure~\ref{fig2}\reb{e} shows clear oscillations with a frequency spacing around $\Delta f = 80$ GHz, where the frequency spacing of the oscillations matches the constructive ($f_\text{max}$) and destructive ($f_\text{min}$) interference conditions at the graphene position, due to the internal cavity, in the near-infrared (NIR) regime. This is given by $\Delta f= \abs{f_\text{max} - f_\text{min}} = c/4 n_{\rm NIR}L$, where $c$ is the speed of light and $n_{\rm NIR}$ the refractive index of silicon in the near-infrared. This gives a refractive index of 3.3, which is close to the reported values~\cite{franta_temperature-dependent_2017} (see Supp.~Fig.~\ref{S_telecom} for details). This shows that the internal cavity leads to interference effects over a broad range of frequencies, spanning at least from 80 GHz to 200 THz. 
\\

\subsection*{Enhanced sensitivity of combined THz detector-interferometer}\label{sec3}

\begin{figure*}[!htp]
    \centering
     \includegraphics[width=\linewidth]{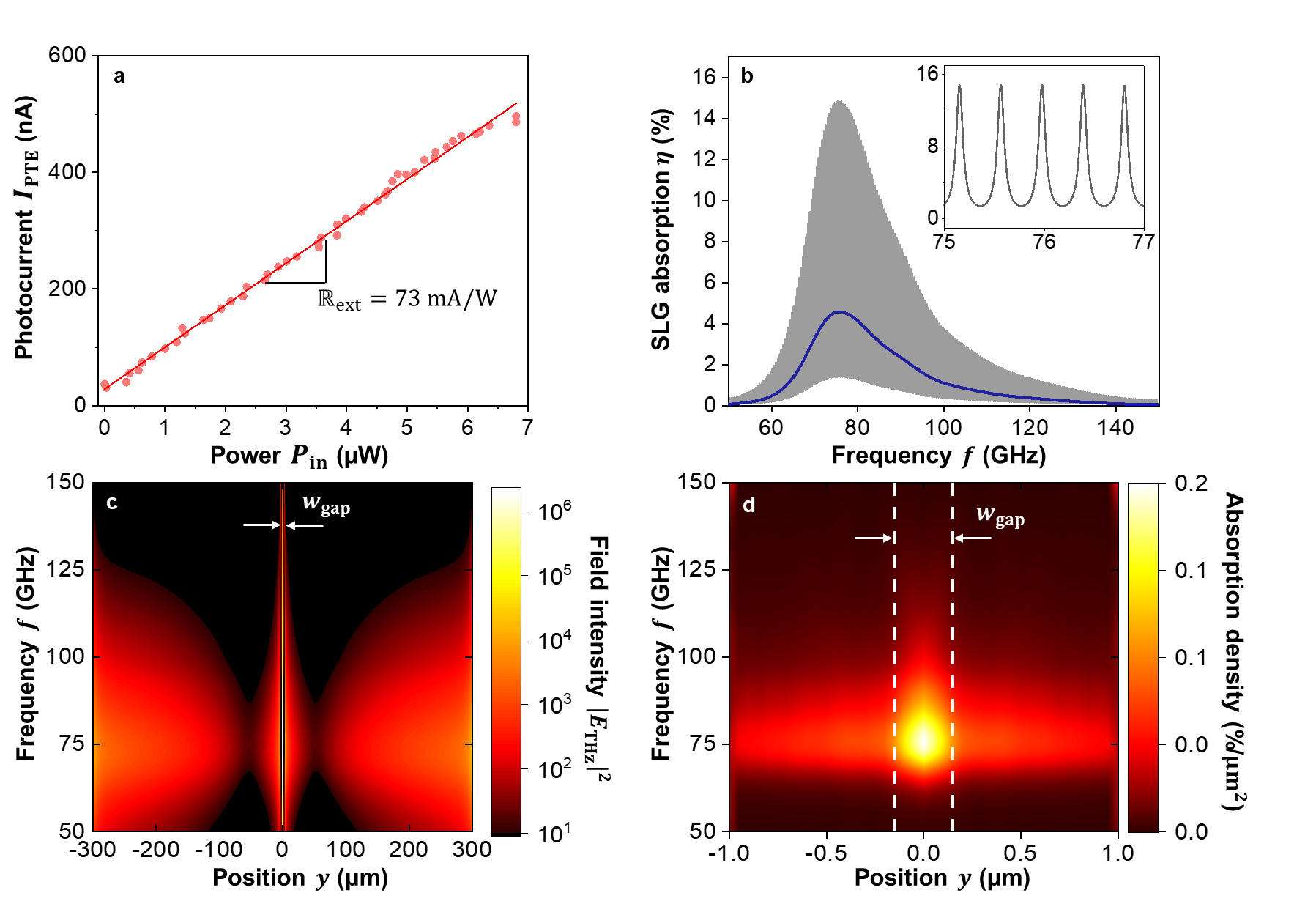}
    \caption{\textbf{Enhanced responsivity and absorption from experiment and simulation}. \textbf{a,} Measured photocurrent $I_\text{PTE}$ at 89 GHz \textit{vs.} incoming THz power $P_{\rm in}$, giving an external responsivity of 73 mA/W. \textbf{b,} Simulation of the \rev{(sub-)}THz absorption $\eta$ \textit{vs.} frequency for incident THz light that is focused with a numerical aperture of 0.5, as in the experiment. The absorption gives a peak value due to interference in the internal cavity (blue curve), in agreement with the experimental results of Fig.~\ref{fig2}\reb{a}. Additional enhancement occurs due to the external cavity (grey curve), which gives rise to a substructure that is shown in the inset. \textbf{c,d,} Maps of the electric field intensity $\abs{E_\text{THz}}^2$ (\textbf{c}) and absorption density (\textbf{d}) as a function of frequency and lateral position $y$, along the source-drain direction. These maps show strong enhancement inside the gap of the dipolar antenna around the resonance frequency. The dashed lines indicate the gap of the dipolar antenna with width $w_{\rm gap}$.}    
    \label{fig3}
\end{figure*}

Having reported evidence of interference occurring in our \rev{(sub-)}THz detector due to a built-in optical cavity, we explore how this enhances the performance of the detector in terms of responsivity and noise-equivalent power. Figure~\ref{fig3}\reb{a} shows the measured photocurrent \textit{vs.} the incident THz power $P_{\rm in}$. The slope gives an external device responsivity of $\mathbb{R}_\text{ext} = I_\text{PTE}/P_\text{in}$ = 73 mA/W (36 V/W). Considering Johnson noise as the main noise source, this gives an external noise-equivalent power (NEP$_{\rm ext}$) of 79~pW$~\rm{Hz}^{-1/2}$ \rev{for Device 1} (see \meths). \rev{Device 4 gives a responsivity of 172 mA/W and an NEP of 26 pW$~\rm{Hz}^{-1/2}$, which we ascribe to a more optimal graphene channel geometry with a width of 3 $\mu$m instead of 30 $\mu$m (see Supp. Figs. S12-13).} These numbers are not corrected for the amount of THz absorption or the ideal case of a diffraction-limited spot size, which means that they are truly external performance parameters without any normalization. In terms of external responsivity and NEP, our detector outperforms all bias-free, graphene-based photodetectors in the \rev{(sub-)}THz regime published to date (see
Supp.~Fig.~\ref{S_table}). 
\\

The large responsivity of the detector could be due to a long cooling time $\tau_\text{cool}$ of hot carriers in graphene, or due to strong THz absorption. To explore the first hypothesis, we determine the cooling time by performing time-resolved photocurrent measurements (see~\meths, and Supp.~Fig.~\ref{S_TRPC}). 
We extract a mean cooling time around 3.4 ps, which is a typical value for hot carrier cooling in hBN-encapsulated graphene~\cite{massicotte_hot_2021,tielrooij_out--plane_2018}. From the experimentally determined photocurrent and cooling time, we estimate a THz absorption coefficient $\eta$ of 20\% for our THz focus spot (see~\meths). This large absorption is the combined result of the antenna, the internal cavity, and the external cavity, as we show in the next section. Importantly, this is the absorption for the case of the experimental, non-diffraction-limited focus area $A_{\rm focus}$. It is common practice to extrapolate the performance of (THz) photodetectors to the situation for a diffraction-limited spot with area $A_{\rm diff}$ by multiplying the responsivity by $A_{\rm focus}/A_{\rm diff}\approx 13$. Interestingly, our detector already absorbs so strongly that such an extrapolation would result in a nonphysical absorption above 100\%. 
\\

\subsection*{Simulations of absorption and responsivity}

To support our experimental findings, we simulate the photoresponse of our device using finite-difference time-domain simulations (see~\meths~for details). We simulate the spectral response in terms of absorption in the graphene layer and the device responsivity in two cases: \textit{i}) considering the internal cavity only, and \textit{ii}) including both the internal and external cavities. The results in Fig.~\ref{fig3}\reb{b} show that the internal cavity increases the graphene \rev{(sub-)}THz absorption by one order of magnitude (see also Supp.~Fig.~\ref{S_simus}). By adding the external cavity, the total absorption in the graphene channel increases by another factor of three, reaching a maximum of 15\%. This is very close to the 20\% obtained from the measured photocurrent and cooling time. Furthermore, we extract $\mathbb{R}_\text{ext} = $ 24 mA/W from the simulation with only the internal cavity, and $\mathbb{R}_\text{ext} = $ 120 mA/W by including internal and external cavities, as shown in Supp.~Fig.~\ref{S_simus}. This is close to the experimentally obtained value of $\mathbb{R}_\text{ext}$ of 73 mA/W \rev{for Device 1}.
\\

The inset of Fig.~\ref{fig3}\reb{b} shows the effect of the external cavity formed between the THz source and the detector-interferometer, which is the appearance of additional maxima and minima in absorption. The periodicity of this substructure is determined by the source-detector distance. We observe this substructure also in the experimental photocurrent data, as shown in the inset of Fig.~\ref{fig2}\reb{a}. Indeed, Fourier analysis of the experimental photocurrent spectrum for two different source-detector distances reveals the periodicity of the additional peaks around the resonant frequency. This periodicity matches the spectral substructure found in the simulations (see Supp.~Fig.~\ref{S_simus}). \rev{Thus, the interference in the internal cavity mainly leads to enhanced responsivity for (sub)THz radiation, while the external cavity mainly gives rise to the observed interference pattern.}
\\

In addition to internal and external cavities, the device also contains a dipolar antenna that enhances the absorption in the graphene channel, as demonstrated earlier~\cite{castilla_fast_2019}. To observe its effect, we simulate the \rev{(sub-)}THz field intensity and absorption density along the graphene channel at different frequencies, see Figs.~\ref{fig3}\reb{c,d}. The distribution of the \rev{(sub-)}THz field intensity on the device surface demonstrates the funnelling of THz light into the antenna gap. Because the graphene channel is located directly below this gap, this leads to strong \rev{(sub-)}THz absorption enhancement and therefore to a large device responsivity. In Supp.~Fig.~\ref{S_simus}, we summarize the relative contributions to the absorption enhancement from the antenna, the internal cavity, and the external cavity. \\

\subsection*{\rev{Quantifying the phase sensitivity}}\label{sec6}

\rev{Having established that the (sub-)THz detector has a high sensitivity and acts as an interferometer, we now quantify its phase sensitivity. We first describe the obtained interference patterns quantitatively, for which we use an Airy function that is typical for a Fabry-P\'{e}rot cavity (see Supp.~Info.~Section \ref{PhaseSensitivity} for details). This gives good agreement for the experimentally obtained and simulated interference patterns, with coefficients of finesse of $\sim$0.05 and $\sim$2, respectively (see Supp.~Fig.~S\ref{suppl_fig28}). We next determine a spatial sensitivity $\partial I_\text{PTE}/\partial x$ of 370 nA/mm by performing the numerical derivative in Figure 2c. Since the phase $\varphi$ is related to the spatial position $x$ via $\varphi=2\pi x/\lambda_0$, the phase sensitivity follows from $\frac{\partial I_\text{PTE}}{\partial \varphi} =\frac{\partial I_\text{PTE}}{\partial x}\cdot\frac{\partial x}{\partial \varphi}$ = 200 nA/rad (see Supp.~Info.~Section \ref{PhaseSensitivity} for details). We define the zero-point of the phase as the point halfway between a peak and a dip in the interference pattern, which is the point with the largest phase sensitivity. Using this definition, we can distinguish phases between $-\pi/4$ and $+\pi/4$.}
\\

\rev{To determine the minimum detectable phase $\delta \varphi$, we use $\delta \varphi=(\partial I_\text{PTE}/\partial \varphi)^{-1}I_\text{noise}$, where $I_\text{noise}$ is the noise current.  We experimentally determine this to be 3.6 nA (see Supp.~Fig.~S\ref{suppl_fig20c}). This gives a phase sensitivity of $\delta \varphi$ = 0.018 rad (= 1$^{\circ}$). If the noise would be limited by the detector, we would obtain a Johnson noise current of 4 pA, which would lead to a phase accuracy of 0.02 mrad. We don’t achieve this, because we are limited by the frequency stability of the source, rather than the detector noise. The reason is that the phase and frequency of the THz source are related to each other via the external cavity via $\varphi=\pi L f/c$, where $L$ is the distance between the THz source and the detector~\cite{Roggenbuck_2010}. In order to quantify the frequency stability, we measure the photocurrent while sweeping the frequency in a small frequency region at different times (see Supp.~Fig.~S\ref{suppl_fig30}) and obtain frequency fluctuations of $\approx$5 MHz. This corresponds to phase noise of $\approx$0.02 rad, similar to the experimentally obtained one. This confirms that the minimum detectable phase is currently limited by the stability of the THz source. Importantly, our approach forms a complete, traceable and uncertainty-bounded framework for phase determination.}
\\

\subsection*{Using phase sensitivity for thicknesses determination}\label{sec4}

\begin{figure*}[!htp]
    \centering
    \includegraphics[width=\linewidth]{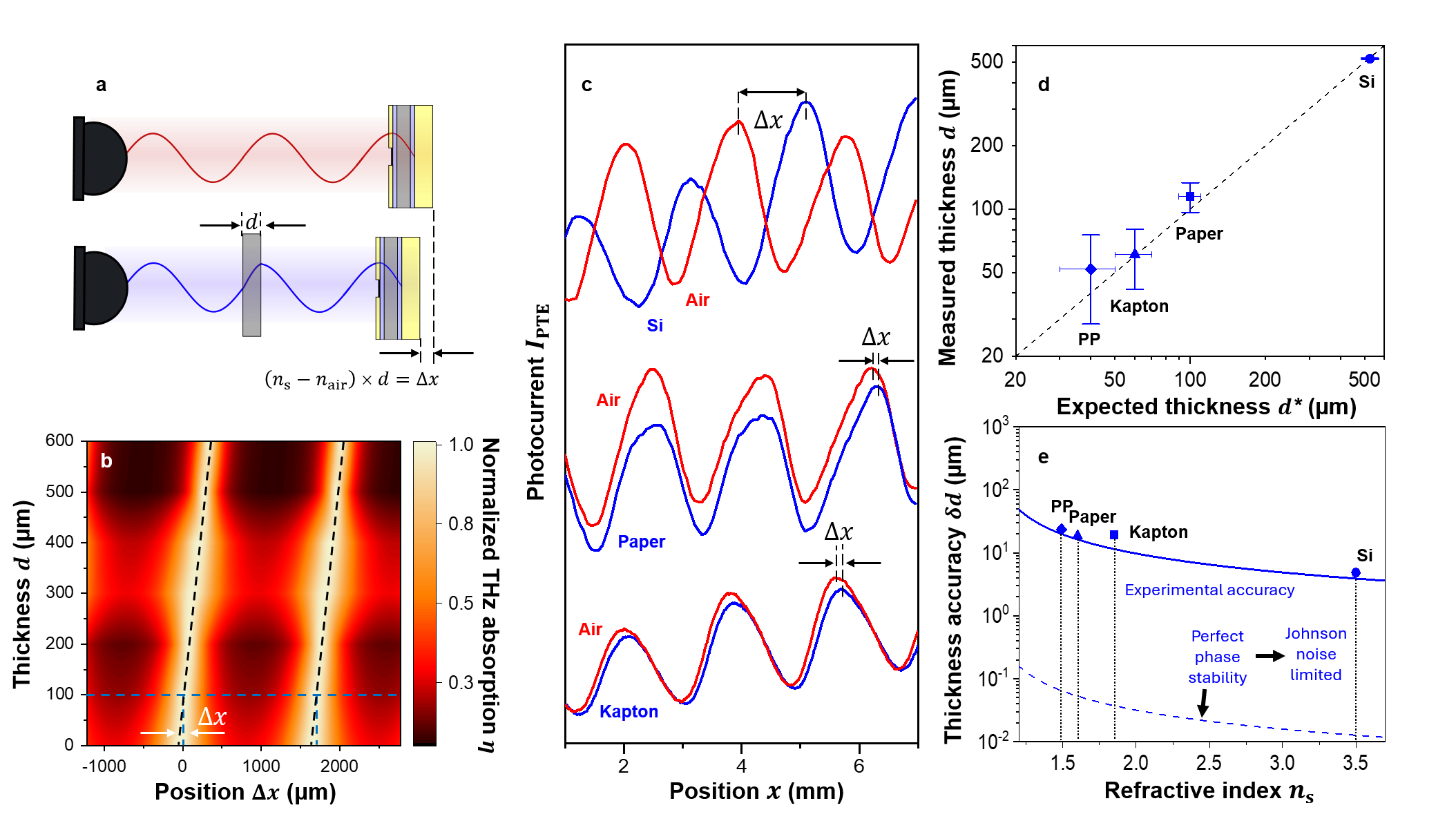}
    \caption{\textbf{\rev{Updated figure panels d and e} Thickness determination using coherent THz detector-interferometer.} \textbf{a,} Scheme of the THz thickness determination measurements, where the spatial shift $\Delta x$ is related to the thickness $d$ of the material placed between the THz source (left) and detector (right). \textbf{b,} Simulation of the absorption in the graphene channel, while moving the detector along the direction of the wave propagation, for the case that a material with $n_\text{s}=1.6$ is placed between the THz source and the detector. The black dashed line shows the maximum absorption of the graphene depending on the thickness of the material $d$. The slope is equal to $(n_\text{s}-n_\text{air})$. The intersection between the blue and black dashed line represents the position of maximum absorption for a thickness of 100 µm. The spatial shift $\Delta x$ is the distance between the position of the intersection and the position where there is no sample, \textit{i.e.}\ $d=0$.  \textbf{c,} Photocurrent $I_\text{PTE}$ interference patterns ($y$ axis offset for clarity) \textit{vs.} position $x$, measured for three combinations: thin films of silicon, paper and kapton (blue), each combined with a reference measurement without sample (red). The spatial shift $\Delta x$ in the interference pattern is the result of the phase delay induced by the higher refractive indices of the samples compared to air. \textbf{d,} Extracted thicknesses $d$ for silicon, paper, kapton, and polypropylene (PP) films (blue markers) \textit{vs.} expected thicknesses $d^*$. The error bars represent the 68\% confidence interval. \textbf{e,} Determination of the thickness accuracy $\delta d$ \textit{vs.} refractive index $n_\text{s}$ in the experimental case (solid line). The dashed line shows the ideal thickness accuracy for a phase-stable source, where the photocurrent noise is only limited by Johnson noise.}
    \label{fig4}
\end{figure*}

\rev{Finally}, we exploit the phase sensitivity to determine the thickness of thin films. In order to do so, we measure the generated photocurrent while moving the detector along the out-of-plane direction $x$, with and without a thin sample material placed between the THz source and our \rev{(sub-)}THz detector-interferometer, as shown in Fig.~\ref{fig4}\reb{a}. Since the sample material has a \rev{(sub-)}THz refractive index $n_\text{s}$, this gives rise to a spatial shift $\Delta x$ in the interference pattern. From this spatial shift, we extract the thickness of the sample $d$ using: $\Delta x = (n_\text{s}-n_\text{air})d$ with $n_\text{air}=1$.
\\

We confirm the validity of this approach by simulating the absorption in graphene when moving along the $x$-position. We use the case where a material of thickness $d$ and refractive index $n_\text{s}=1.6$ (for paper) is placed between the source and the detector. The results in Fig.~\ref{fig4}\reb{b} show that maximum absorption occurs for specific combinations of thickness $d$ and position $x$. These maxima occur with a periodicity of $\lambda_0/2$ and satisfy the relation $\frac{\partial x}{\partial d} = (n_\text{s}-n_\text{air})$. This simulation confirms that our experimental approach to determine sample thicknesses is valid.  
\\

Figure~\ref{fig4}\reb{c} shows the results of the phase-enabled thickness measurements on three different materials widely used in industrial sectors: a high-resistive double-side polished Si wafer (Si, $n_\text{s} \approx 3.5$~\cite{franta_temperature-dependent_2017}), paper ($n_\text{s} \approx 1.6$~\cite{Zhai2021-ms}), and a Kapton polyimide thin film ($n_\text{s} \approx 1.8$~\cite{cunningham_broadband_2011}). In all cases, the oscillating pattern is displaced by a spatial shift $\Delta x$ between the case without and the case with the sample. The shift depends on the thickness and refractive index of the inset material. The thicknesses we obtain are consistent with the independently determined thickness values, as shown in Fig.~\ref{fig4}\reb{d} (see~\meths~for details). This also includes a measurement of a polypropylene (PP) film with a thickness of 40 µm and refractive index of 1.5~\cite{sahin_dielectric_2019}(see Supp.~Fig.~\ref{S_imaging}). 
\\

The precision of the thickness measurement depends on how much the photocurrent changes due to a change in the thickness of the material with a certain refractive index, compared to the smallest observable change in the photocurrent: $\delta d = \left( \frac{\partial I_\text{PTE}}{\partial x} \right)^{-1} \times \frac{I_\text{noise}}{(n_\text{s}-1)}$. Figure~\ref{fig4}\reb{d} shows the calculated thickness accuracy as a function of the refractive index, which closely matches the thickness error bars that we obtained experimentally for the four different materials that we studied. We obtain the highest accuracy for silicon, which is a few micrometers. \rev{We haven taken the refractive index of the samples as given. In case of uncertainty in the refractive index $\delta n_s$, the error will propagate as $\delta d/d= \delta n_s/(n_s - 1)$.}
\\

If the photocurrent noise were limited by Johnson noise, we would reach a thickness accuracy close to 10 nm, which is six orders of magnitude smaller than the wavelength. This requires a \rev{sufficiently} phase- \rev{and power-}stable THz source, faster scanning of the detector position, and \rev{potentially} packaging of the detector to minimize electronic noise. A more intense THz source \rev{could} lead to a further improvement in thickness accuracy.
\\

\section*{Discussion}\label{sec5}

In conclusion, we have added the capability of phase sensitivity to \rev{(sub-)}THz detectors by developing a chip-scale, integrated \rev{(sub-)}THz detector-interferometer with high responsivity and high sensitivity to the phase of the incident \rev{(sub-)}THz light. This is enabled by combining the strong photo-thermoelectric response of graphene with a planar THz dipolar antenna and a vertical optical cavity. We have demonstrated interference effects in the spectral, spatial, and temporal domain, and for both \rev{(sub-)}THz and near-infrared light. Finally, we have exploited the phase sensitivity to determine the thickness of thin films with deep sub-wavelength accuracy. \rev{ The main advantages of our integrated detector-integrator concept are i) its very low power consumption, as the device is passive, in particular not requiring any bias voltage; ii) its compactness; iii) its operation with continuous-wave light, rather than femtosecond pulsed light; and iv) its broadband operation that is only limited by the antenna and the internal cavity, and not by the photo-active material.} This makes the graphene-based detector-interferometer a promising alternative to THz time-of-flight measurements using femtosecond pulses, aimed at determining film thicknesses, for example. 
\\

While our optical cavity led to the largest absorption enhancement and phase sensitivity for light around 80-90 GHz, it is possible to optimize the optical cavity for other frequencies, or to incorporate a tunable cavity~\cite{yao_configurable_2024}. \rev{The cavity resonances also repeat for any integer of wavelengths in the cavity. Replacing the dipolar antenna by a broadband antenna, such as a bow-tie or log-spiral antenna, also broadens the applicable wavelength range. Finally, we note that for several applications only a single-pixel detector is required, which means that scalable device fabrication, for example via growth based on chemical vapor deposition, is not a strict prerequisite for applicability.} Beyond thickness measurements, the phase sensitivity of our \rev{(sub-)}THz detector-interferometer opens up a myriad of possibilities to exploit phase information, including non-destructive industrial testing, beyond-fifth-generation wireless data transfer, and quantum technologies.
\\

\backmatter

\section*{Materials and methods}
\label{meth}

\textbf{Device fabrication.} The starting point of the fabrication \rev{of Device 1} was a highly resistive double-side polished silicon substrate with a thickness of 279 µm, with thermally grown SiO$_2$ of thickness 300 nm on both sides. We placed hBN-encapsulated graphene on top, and contacted the graphene to source and drain electrodes via side contacts using etching and metal evaporation. The graphene channel has a width of $w_\text{dev} = 30$ µm. We deposited a dipolar antenna on top of the hBN-graphene-hBN stack. The antenna has a gap width $w_\text{gap}$ of 300 nm, a width $w_\text{ant}$ of 30 µm, and a length $L_\text{ant}$ of 681 µm. The substrate with graphene, contacts, and antenna was placed on a metal support, forming the back mirror. \rev{We note that Device 1 has been measured over the course of three years and across three different countries with good reproducibility.} \rev{Devices 2-4 were fabricated using a polished silicon substrate with a thickness of 525 $\mu$m, with a 100 nm layer of Al$_2$O$_3$ grown using atomic layer deposition. We first spin-coated a photolithography resist to protect the top side of the wafer, then we evaporated 5 nm of Ti and 50 nm of Au on the back side of the wafer to create the back metallic mirror. We next dissolved the photoresist using acetone and performed a post-clean process using oxygen plasma. We placed hBN-encapsulated graphene on the top side of the wafer and contacted the graphene as we described for Device 1. We used a channel length of 1 $\mu$m for Devices 2-4, and a width of 1 $\mu$m for Devices 2-3 and 3 $\mu$m for Device 4. We next placed a layer of 30 nm of hBN using dry transfer technique, and then we patterned a bow-tie antenna on top with a gap of 300 nm using electron beam lithography. }
\\

\textbf{Terahertz photocurrent setup.} The terahertz photocurrent setup is based on a ``Terascan 1550'' continuous wave terahertz spectroscopy system~\cite{deninger_275_2015} (Toptica Photonics AG), which has an accessible range from 0 to 1.2 THz, with a THz power around 7.6 µW at 89 GHz. Power calibration curves are available in Supp.~Fig.~\ref{S_methods}. The divergent THz beam generated on the source side of the commercial THz setup is collimated using a first parabolic mirror and then focused onto the graphene-based detector-interferometer by a second parabolic mirror. Both parabolic mirrors have a focal distance of 50.8 mm and a diameter of 50.8 mm, corresponding to a numerical aperture (NA) of 0.5. The sample is mounted on a motorized $xyz$-stage, whose maximum range along one axis is 25 mm and minimal step resolution of 1.25 µm. We collect the demodulated photocurrent via a lock-in amplifier (Zurich Instruments MFLI). The reference frequency to perform the lock-in detection is based on the AC bias applied to the THz source (39.7 kHz). All measurements are performed with $V_g^\text{A} = V_g^\text{B} = 0$ V. For the THz thickness measurements, the samples of interest (Si wafer, paper, Kapton, polypropylene film) are placed close to the THz focal point in front of the graphene-based detector-interferometer. In all measurements, a polystyrene foam is placed in the collimated beam path to absorb any remaining non-THz light that was used to generate THz radiation. An illustration of the setup is depicted in Supp.~Fig.~\ref{S_methods}.
\\

\textbf{Time-resolved photocurrent setup.} We use a femtosecond laser ($\simeq$100 fs pulse duration, 76 MHz repetition rate, 1030 nm central wavelength) split into two pump beams to excite the device. We modulate one of these pump beams at 240 Hz using a mechanical chopper. This beam passes through a mechanical delay line that controls the time delay $\tau_\text{delay}$ between both pump pulses. For both pump beams, we adjust the power at the sample location to a few hundreds of µW. We set the polarization of the pump beams perpendicular to each other. We combine both beams using a non-polarizing beam splitter, and a 50x objective (NA $=0.42$) focuses the beams onto the photo-active region of the graphene-based detector-interferometer, with typical spot sizes of $\simeq$1 µm. We use closed-loop $xyz$-nanopositioners to move the sample with respect to the laser beams. We collect the generated photocurrent from the device using lock-in detection by demodulating the signal at the chopper frequency. Additionally, we can collect the reflected probe light using a biased InGaAs photodetector to simultaneously obtain reflectance and photocurrent maps of the device. For the time-resolved photocurrent measurements, we scan the delay line to achieve a time delay window of $\pm$30 ps, with sub-ps temporal resolution. We perform all measurements under ambient conditions without applying any gate voltage to the device. An illustration of the setup is shown in Supp.~Fig.~\ref{S_methods}. To extract the cooling time, we fit the photocurrent signal as a function of delay time to exponentially decaying functions. 
\\

\textbf{Photoresponse simulations.} We perform all the simulations using Lumerical software. We first simulate the frequency response of the cavity using a 2D model with a plane wave for three cases: \textit{i}) without a back mirror, \textit{ii}) with back mirror, \textit{iii}) including a Si lens in the point source. In the three cases, we change the distances of the point source with respect to the device to observe the cavity effects induced by the source and the \rev{(sub-)}THz cavity. Then, we perform a simulation using a 3D model and use a focused THz beam as a source to determine the absorption on the graphene. More details are available in Supp.~Fig.~\ref{S_simus}. The descriptions of the thermoelectrical and electrostatic models are available in Refs.~\cite{vangelidis_unbiased_2022} and~\cite{castilla_electrical_2024}.  \\

\textbf{Obtaining the detector responsivity and NEP.} The external responsivity for our experimental THz spot size is defined by $\mathbb{R}_\text{ext} = I_\text{PTE}/P_\text{in}$, where $I_\text{PTE}$ is the peak-to-peak value of the demodulated PTE photocurrent signal measured with the lock-in amplifier and $P_\text{in}$ is the incoming power measured at the focus position. We calculate the peak-to-peak value of the photocurrent using $I_\text{PTE}=2\sqrt{2}\pi/4\cdot I_\text{lock-in}$, where $I_\text{lock-in}$ is the root mean square value of the demodulated photocurrent signal measured by the lock-in amplifier. The external noise-equivalent power is $\text{NEP}_\text{ext} = I_\text{noise}/\mathbb{R}_\text{ext}$, where $I_\text{noise} = \sqrt{4 k_\text{B} T \Delta f/ R}$ is the Johnson noise, $k_\text{B}$ is the Boltzmann constant, $T = 300$~K is the temperature, $\Delta f = 1$~Hz is the measurement bandwidth and $R = 500~\Omega$ is the channel resistance. In the experiment, the non-diffraction-limited focus area was around 57.8 mm$^2$ (see Supp.~Fig.~\ref{S_methods}). The ideal diffraction-limited focus area is given by~\cite{castilla_fast_2019} $A_{\rm diff} = \frac{\lambda_0^2}{\pi} \approx$ 4.5 mm$^2$. This gives a ratio $A_{\rm focus}/A_{\rm diff}\approx$ 13. 
Since the absorption is already 15-20\% in the non-diffraction-limited case, we can expect 100\% absorption for a (near-)diffraction-limited spot, giving an ideal responsivity $\mathbb{R}_{\rm diff}$ (ideal NEP$_{\rm diff}$) that is increased (decreased) by at least a factor five.
\\

\textbf{Obtaining the absorption from experiments.} We first extract the increase in temperature from the photocurrent using $I_\text{PTE} = (S_1-S_2)\Delta T/R$~\cite{castilla_fast_2019}, where $S_1-S_2$ is the Seebeck coefficient (160 µV K$^{-1}$, known from previous reports~\cite{soundarapandian_high-speed_2024,castilla_fast_2019}), and $R=500~\Omega$ is the resistance of the device. This gives $\Delta T$ = 1.55 K for the highest THz power. The following equation describes the relationship between the increase in temperature and the absorption $\eta$: $ \Delta T= \frac{\eta P_\text{in} \tau_\text{cool}}{2l_{\text{cool}} w C_{\rm el}}$, where $C_{\rm el}=1 \times 10^{-4}$ Jm$^{-2}$K$^{-1}$ is the electronic heat capacity, $\tau_\text{cool}$ is the cooling time of graphene, $l_\text{cool}=500$ nm is the cooling length in graphene and $w=30$ $\mu$m the width of the photoactive area. The cooling length is estimated from the cooling time and the momentum relaxation time~\cite{massicotte_hot_2021}. Using these parameters, which are all known, we extract the absorption coefficient. 
\\

\textbf{Thickness measurements.} \rev{We first measured the thickness of all the materials using a commercial caliper with an accuracy of 10 $\mu$m as a reference value for comparison. Then, we measured the thickness of all the materials by performing a statistical averaging to improve the accuracy of the thickness determination and to simultaneously mitigate the intrinsic phase drift of our THz system.} Concretely, we measure around twenty consecutive scans close to where a local maximum is located (for air and with the material, respectively, see Supp.~Fig.~\ref{S_imaging}).
To determine the peak position of the curve, we find the maximum value of all the curves and extract $\Delta x$ from each consecutive scan (air and material). Then, the thickness $d$ is given by $\Delta x / (n_\text{s}-n_\text{air})$. Finally, we calculate the average thickness and its standard deviation. This procedure was applied to two groups of data \rev{without applying any outlier rejection}: the first group is the acquired data without any treatment, and the second group is the acquired data treated by smoothing (see Supp.~Fig.~\ref{S_imaging} for more details).
\\

\bmhead{Supplementary information}

The Supplementary Information presents details of the fabrication of the device, calculation of interference conditions, optical simulations, telecom-generated photocurrent, comparison with other detectors performances, time-resolved photocurrent measurements, additional measurements aimed at thickness extraction, and details of experimental methods.

\bmhead{Acknowledgments}

The ICN2 is funded by the CERCA programme / Generalitat de Catalunya. The ICN2 is supported by the Severo Ochoa Centres of Excellence programme, Grant CEX2021-001214-S, funded by MCIU/AEI/10.13039.501100011033. K.-J.T. acknowledges funding from the European Union’s Horizon 2020 research and innovation program under Grant Agreement No. 804349 (ERC StG ``CUHL'') and No. 101125457 (ERC CoG ``EQUATE''). S.C., K.P.S. and F.H.L.K acknowledge funding by the European Union program under Grant Agreement No. 101113529 (ERC, TERACOMM) and PDC2022-133844-I00, funded by MCIN/AEI/10.13039/501100011033 and by the “European Union NextGenerationEU/PRTR". K.W. and T.T. acknowledge support from the JSPS KAKENHI (Grant Numbers 21H05233 and 23H02052), the CREST (JPMJCR24A5), JST and World Premier International Research Center Initiative (WPI), MEXT, Japan.

\bmhead{Data availability statement} 

The data that support the findings of this study are available from the corresponding author(s) upon reasonable request.

\bmhead{Competing interests}

The authors declare no competing interests.

\bmhead{Author contributions}

S.C., F.H.L.K. and K.-J.T conceived the project. R.dlB., E.R. and A.N. conducted the experiments. R.dlB., D.S.R and A.N. performed the TRPC measurements. K.P.S. and S.C. fabricated the device. I.V. and E.L. performed the simulations. S.C. and K.P.S. assisted with the measurements and discussion of the results. K.W. and T.T. synthesized the hBN crystals. R.dlB., E.R., A.N., S.C., F.H.L.K. and K.-J.T. wrote the manuscript with input from all the authors. S.C., F.H.L.K. and K.-J.T. supervised the work and discussed the results. All authors contributed to the scientific discussions and manuscript revisions.

\clearpage
\bibliography{sn-bibliography}

\end{document}


\maketitle

The Supplementary Information contains:
\begin{itemize}
    \item \ref{S_fab}: \nameref{S_fab}
    \item \ref{S_cavity}: \nameref{S_cavity}
    \item \ref{S_simus}: \nameref{S_simus}
    \item \ref{S_telecom}: \nameref{S_telecom}
    \item \ref{S_dev}: \nameref{S_dev}
    \item \ref{S_table}: \nameref{S_table}
    \item \ref{S_TRPC}: \nameref{S_TRPC}
    \item \ref{S_airy}: \nameref{S_airy}
    \item \ref{S_imaging}: \nameref{S_imaging}
    \item \ref{S_methods}: \nameref{S_methods}
\end{itemize}

\clearpage
\section{Fabrication of the graphene (sub-)THz detector}
\label{S_fab}

\begin{figure}[!htp]
    \centering
    \includegraphics[width=\linewidth]{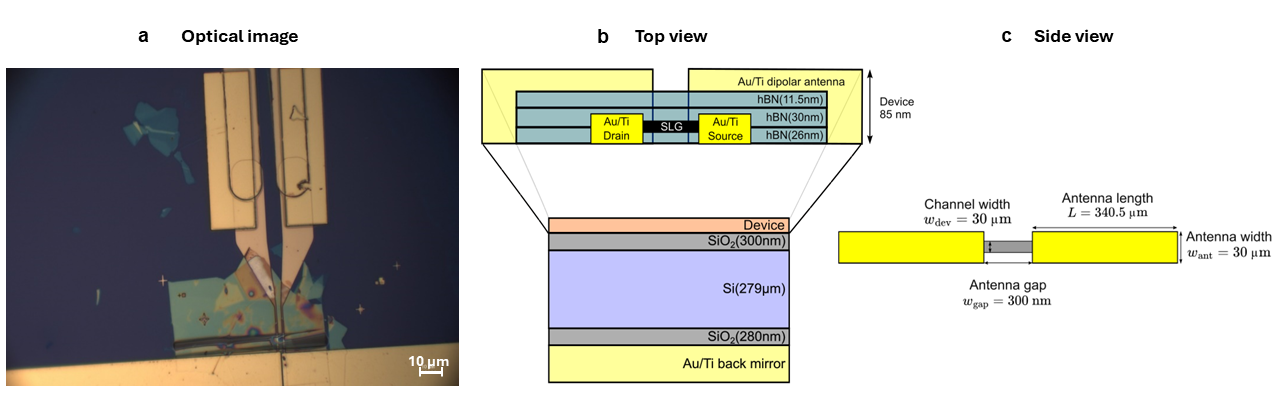}
    \caption{\textbf{THz detector device with integrated cavity.} \textbf{a,} Optical microscope image of the device, where the antenna, source and drain contacts, and the hBN-encapsulated graphene stack are identified. \textbf{b,} Schematic side-view of the device. A metallic plate is engineered at the back of a Si-based (sub-)THz photodetector based on single layer graphene (SLG). In this way, the Si interlayer forms a cavity resonant at 89 GHz. An antenna is engineered on top of the SLG to efficiently couple the incoming THz light with the device. \textbf{c,} Schematic top-view of the device.}
    \label{suppl_fig1}
\end{figure}

\begin{figure}[!htp]
    \centering
    \includegraphics[width=\linewidth]{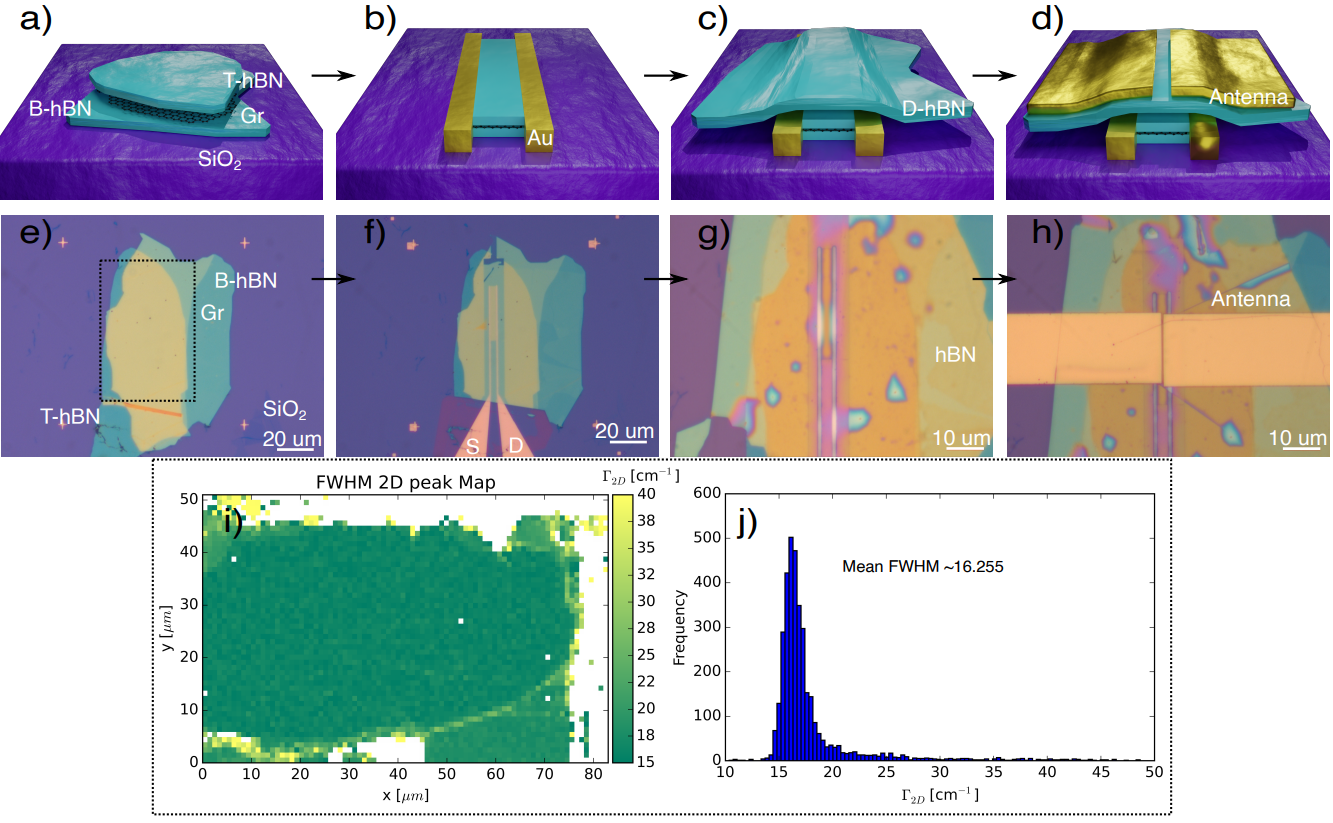}
    \caption{\textbf{Fabrication process.} Step-by-step fabrication procedure including \textbf{(a)} the hBN-encapsulation of the SLG, \textbf{(b)} the etching of the channel and the deposition of the edge contacts, \textbf{(c)} the top insulation with hBN and \textbf{(d)} the deposition of the top antenna. The optical images \textbf{(e)} to \textbf{(h)} respectively present the corresponding fabrication steps \textbf{(a)} to \textbf{(d)}. \textbf{i,} Spatial map of the full width at half maximum (FWHM) of the 2D peak of graphene $\Gamma_\text{2D}$ along with \textbf{(j)} the number of occurrence \textit{vs.} the FWHM of the 2D peak of graphene.}
    \label{suppl_fig2}
\end{figure}

\textbf{Device description.} The structure of the graphene-based photodetector devices with a (sub-)THz cavity is shown in Suppl.~Fig.~\ref{suppl_fig1} and is based on our previous work on photodetectors exploiting the photothermoelectric effect (PTE)~\cite{castilla_fast_2019,soundarapandian_high-speed_2024}. The step-by-step fabrication process is shown in Suppl.~Fig.~\ref{suppl_fig2}, which includes selecting the graphene flake, hBN encapsulation, etching the channel, depositing the source and drain contacts, and depositing the two antenna branches, which simultaneously serve as dual top gates.

\clearpage
\section{Interference conditions in the internal cavity}
\label{S_cavity}

\begin{figure}[!htp]
    \centering
    \includegraphics[width=0.7\linewidth]{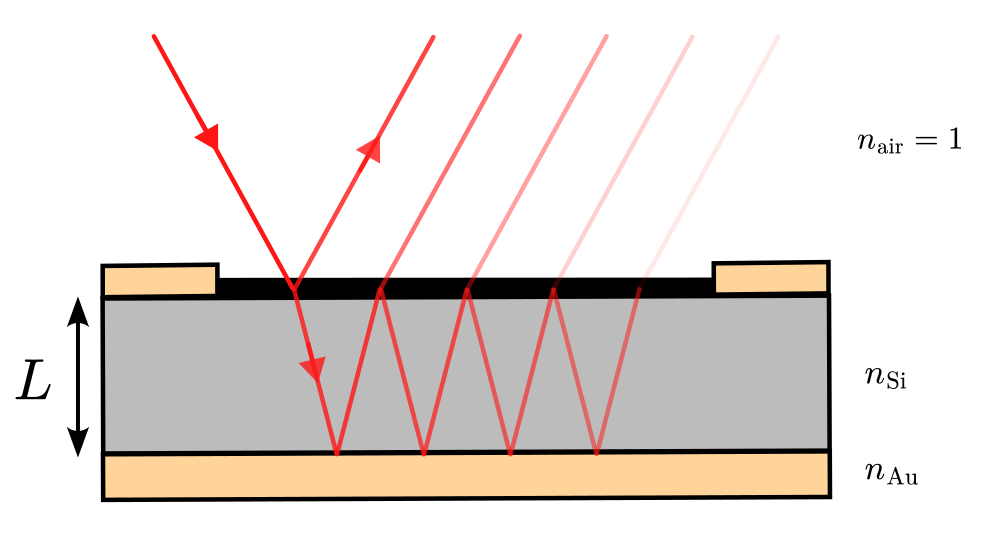}
    \caption{\textbf{Constructive and destructive interferences.} Illustration of the interferences conditions in the internal cavity of thickness $L$ and refractive index $n_\text{Si}$.}
    \label{suppl_fig27}
\end{figure}

\textbf{Derivation of the conditions for constructive and destructive interference.} We derive the interference conditions by considering an incident beam and reflected beams at the position of the graphene channel that serves as photoactive material for the detector, as illustrated in Suppl.~Fig.~\ref{suppl_fig27}. Making a round trip in the cavity brings to the reflected wave a phase term $\phi = 2\times 2\pi n_\text{Si}L/\lambda_0 + \pi$, where the last term comes from the fact that $n_\text{Au} > n_\text{Si}$ \cite{ordal_optical_1987}. The interference conditions for the field intensity are the following:

\begin{equation}
\begin{split}
    &\Delta \phi = 0 +2\pi M \Longrightarrow (2M-1)\lambda_0 = 4n_\text{Si} L~\text{(constructive)}\\
    &\Delta \phi = \pi +\pi M \Longrightarrow 2M \lambda_0 = 4n_\text{Si} L~\text{(destructive)}
    \label{eq_S_condition_condes}
\end{split}
\end{equation}

The lowest frequency that leads to constructive interference at the graphene plane corresponds to $M$ = 1 and gives a resonance frequency of:

\begin{equation}
    f =  \frac{c}{4 L n_\text{Si}}
    \label{eq_S_FB}
\end{equation}

\clearpage
\section{Optical simulations}
\label{S_simus}
\begin{figure}[!htp]
    \centering 
    \includegraphics[width=1\linewidth]{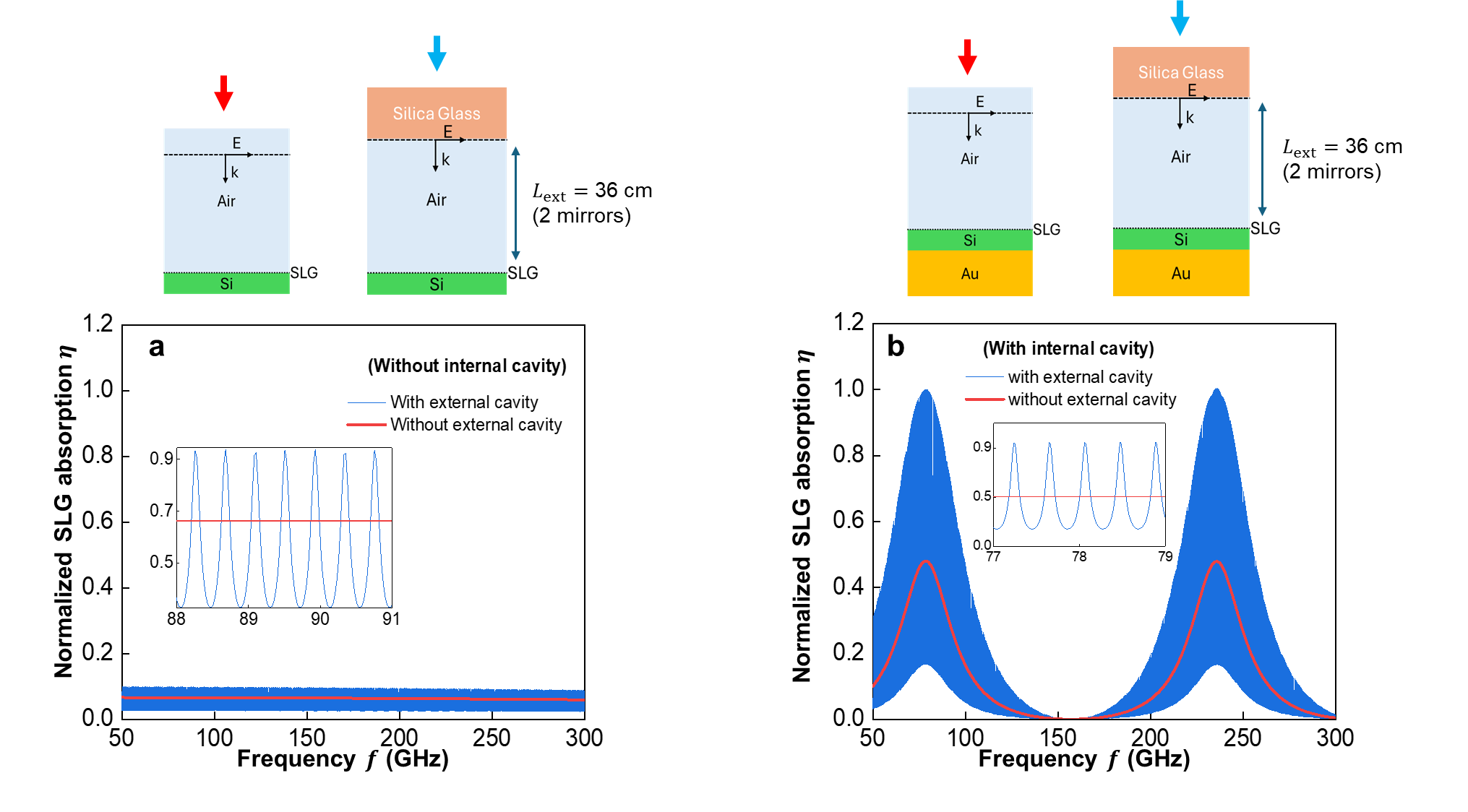}
    \caption{\textbf{Effect of the internal and external (sub-)THz cavities.} \textbf{a,} SLG absorption considering the device without any cavity (left scheme, red curve) and with the external cavity only (right scheme, blue curve). \textbf{b,} SLG absorption considering the effect of the internal cavity (left scheme, red curve) and the effect of both cavities (right scheme, blue curve). In all cases, the introduction of the external cavity corresponds to a source-to-detector distance $L_\text{ext} = 36$~cm, and the values correspond to THz light focused with an NA of 1. The insets in \textbf{(a)} and \textbf{(b)}  show the periodic oscillations induced by the introduction of the external cavity.}
    \label{suppl_fig14}
\end{figure}

\textbf{Simulation of the external cavity.} We simulate in Suppl.~Fig.~\ref{suppl_fig14}\reb{a} the effect of only the external cavity (formed between the SLG and the silica layer representing the emitter surface). We notice a fringe pattern introduced by the external cavity. The magnitude of the enhancement is typically estimated around $\pm$50 \% compared to the simulation without the external cavity. We simulate in Suppl.~Fig.~\ref{suppl_fig14}\reb{b} the effect of the internal cavity, which induces a larger absorption (up to 40\% of absorption, revealing an increase by a factor close to $\times 10$ compared to the reference case without any cavity) in the SLG layer at specific frequencies (78 and 235 GHz). This demonstrates that the internal cavity acts as a (sub-)THz cavity according to the resonance conditions defined in Suppl.~Eq.~\ref{eq_S_FB}. We also observe two times more absorption when simulating the case with both an external and an internal cavity compared to the case with only the internal cavity. \\

\textbf{Simulation of external responsivity in graphene.} We first simulated and calculated the (sub-)THz absorption in the graphene channel in Suppl.~Fig.~\ref{suppl_fig14}\reb{a}, taking into account both the internal cavity (metallic back reflector) and the dipole antenna; only dipole antenna without internal cavity; only internal cavity without dipolar antenna; and without the internal cavity nor dipolar antenna. We observe that the best configuration to increase the absorption is using an internal and external cavity. We next calculate the external responsivity as a function of the frequency for a NA of 0.5 and two different situations : (i) including the internal and external cavity, and (ii) including only internal cavity. We observe that the external cavity enhances the responsivity supporting the previous simulations in Suppl.~Fig.~\ref{suppl_fig14}\reb{b}.\\  

\begin{figure}[!htp]
    \centering
    \includegraphics[width=\linewidth]{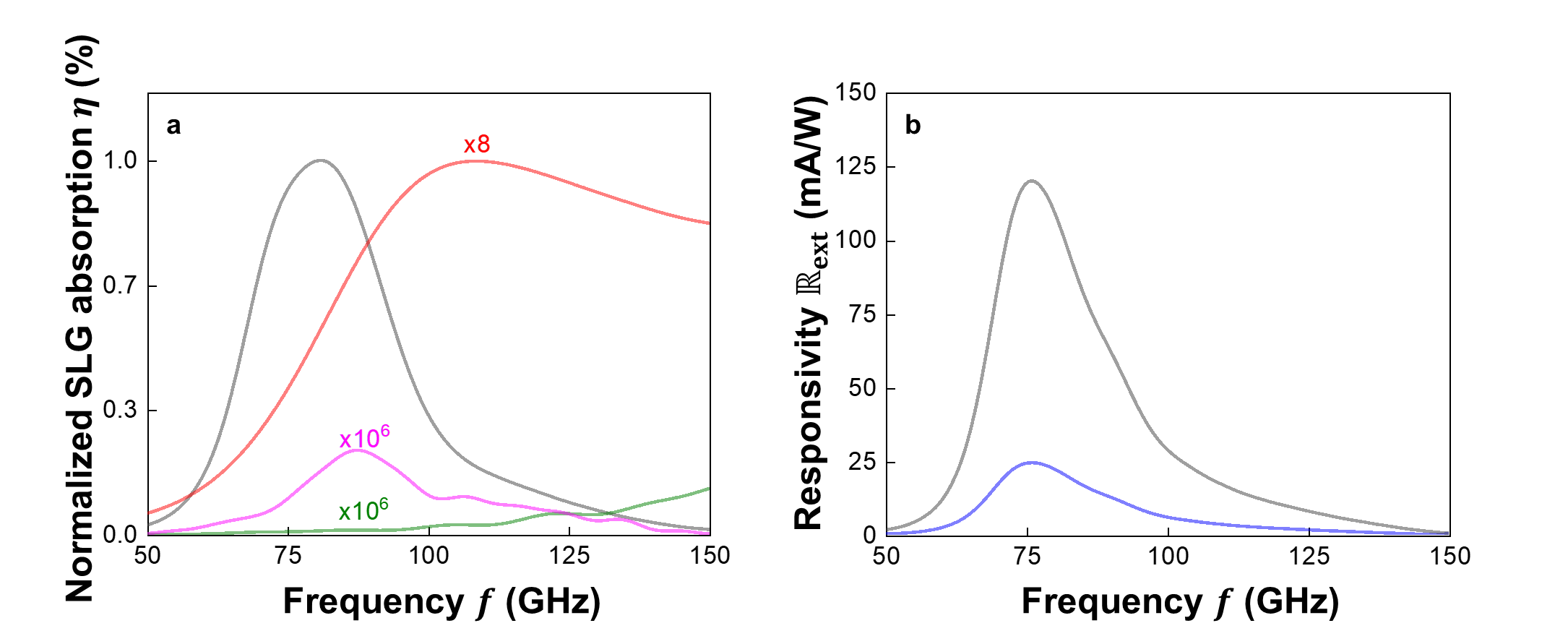}
    \caption{\textbf{Simulation results for absorption and responsivity.} \textbf{a,} Calculated (sub-)THz absorption in the graphene channel, taking into account both the internal cavity (metallic back reflector) and the dipole antenna (grey); only dipole antenna without internal cavity (red); only internal cavity without dipolar antenna (pink); and without the internal cavity nor dipolar antenna (green). We multiply each curve by the factor written above it for better visualization. \textbf{b,} Calculated external responsivity as a function of frequency with (grey) and without (blue) the external cavity.}
    \label{suppl_fig14a}
\end{figure}

\begin{figure}[!htp]
    \centering
    \includegraphics[width=\linewidth]{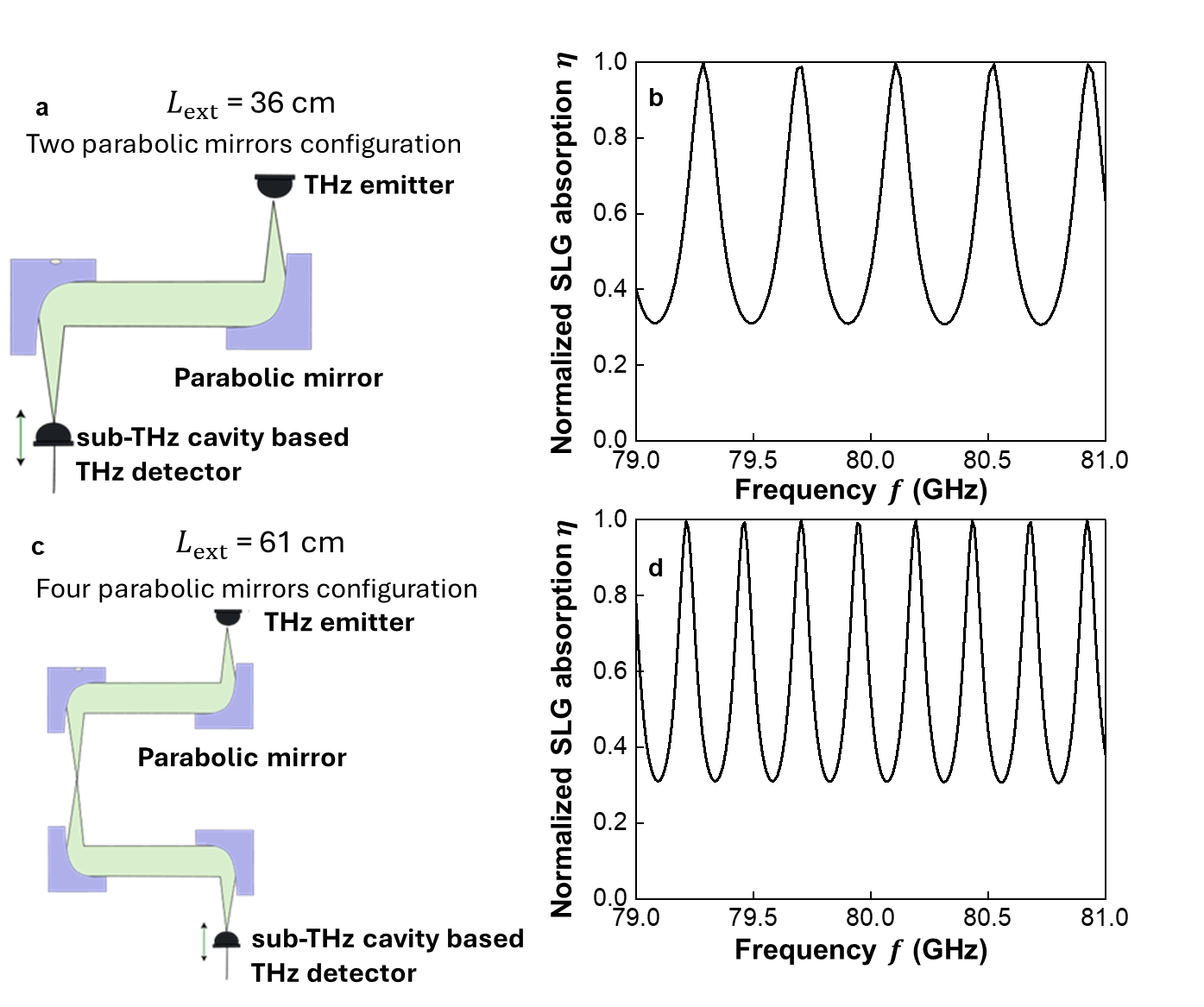}
    \caption{\textbf{Sub-fringes structure in the spectrum induced by external source-detector cavity.} We simulate two cases: \textbf{(a)} two parabolic mirrors configuration where the distance between the emitter and the detector is 36 cm. We observe in \textbf{(b)} the SLG absorption, oscillating with a frequency of $c/2L_\text{ext}$.  For the second case, we simulate a \textbf{(c)} four parabolic mirrors configuration with a distance between the emitter and the detector of 61 cm. We observe in \textbf{(d)} the SLG absorption oscillating with a frequency of $c/2L_\text{ext}$.}
    \label{suppl_fig15}
\end{figure}

\begin{figure}[!htp]
    \centering
    \includegraphics[width=\linewidth]{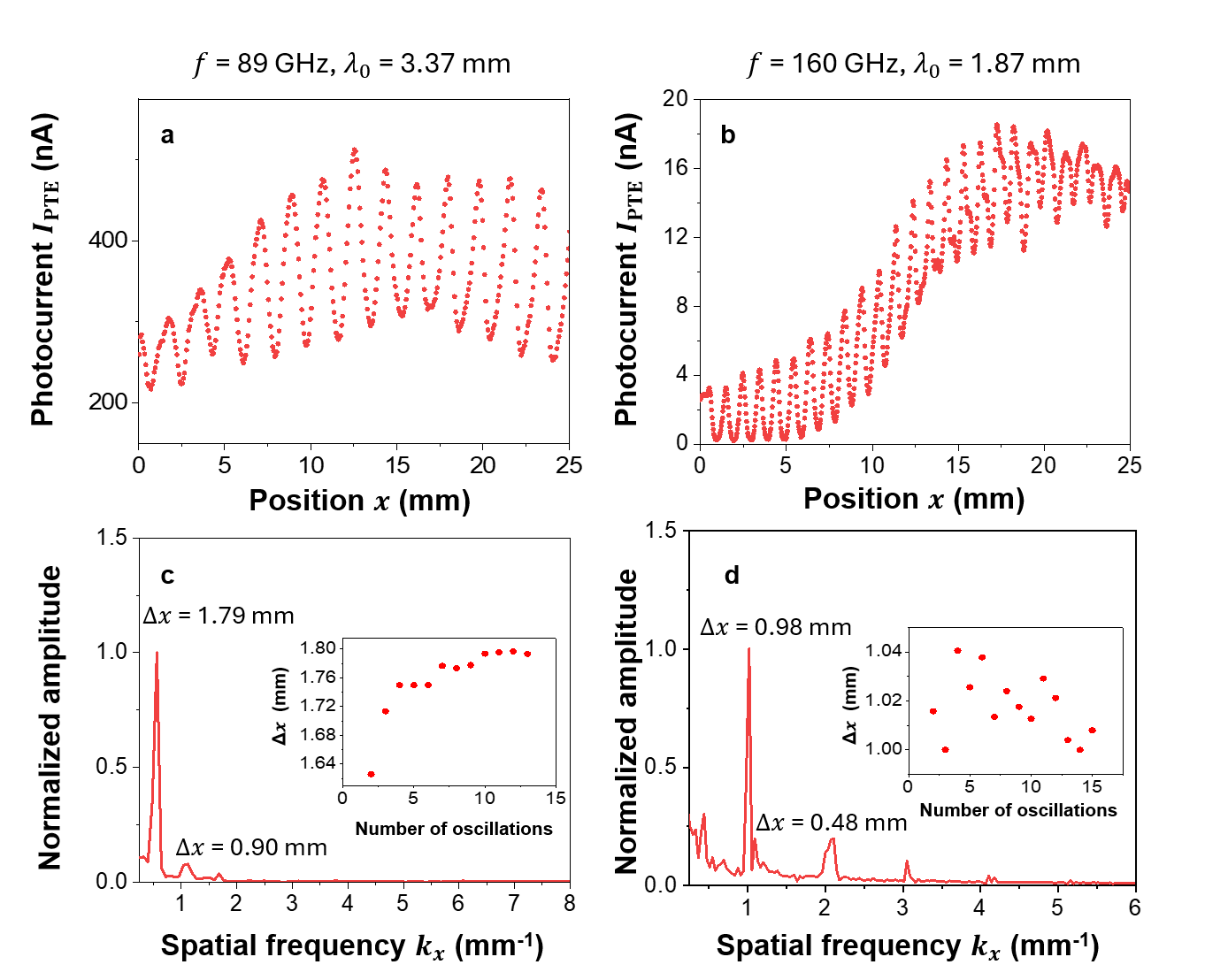}
    \caption{\textbf{Interference pattern observed when scanning the detector position.} \textbf{a,b,} Photocurrent measured while moving the detector along the travel direction of the light for incident radiation at 89 GHz \textbf{(a)} and 160 GHz \textbf{(b)}, together with the Fourier transforms of these interference patterns \textbf{(c,d)}. These results show oscillations due to interference with a period of ~3.35 mm,  on top of an envelope due to focusing, with a length scale of more than 10 mm, as expected for this wavelength and the numerical aperture of 0.5. The Fourier transforms are performed to determine the periodicity of the signals and give main peaks at 1.79 mm and 0.98 mm for 89 and 160 GHz, respectively. This is close to half of the wavelength of the incident light, which is 1.69 mm and 0.94 mm, respectively. The inset in panels (c) and (d) show that the deviation between the fringe period obtained using 2 and 13 cycles of the interference pattern in panel (a) is around 10\%, while the deviation between using 2 and 15 cycles of the interference pattern in panel (b) is around 1\%. This is likely the result of frequency and (associated) phase drift. The fact that we observe oscillations for both frequencies shows that the occurrence of interference is not limited to the resonance frequency of the internal cavity.}
\label{suppl_fig15a}
\end{figure}

\begin{figure}[!htp]
    \centering
    \includegraphics[width=\linewidth]{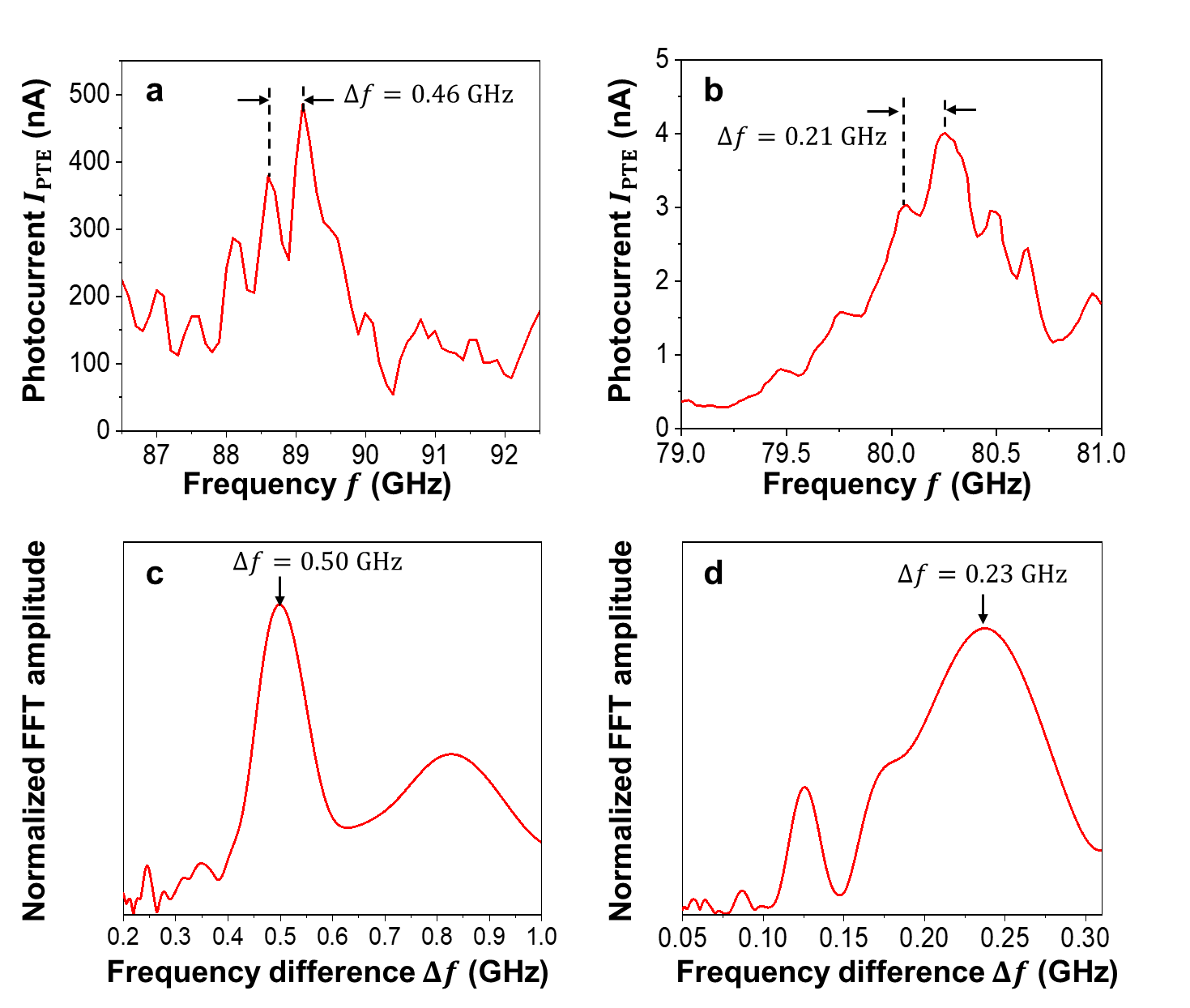}
    \caption{\textbf{Experimental signatures of the external cavity.} 
    \textbf{a,b} Photocurrent signal for source-detector distances $L_\text{ext}$ of 36 cm \textbf{(a)} and 61 cm \textbf{(b)}, showing a spectral substructure. \textbf{c,d} Fourier transforms of the photocurrent signals for $L_\text{ext}$ of 36 cm \textbf{(c)} and 61 cm \textbf{(d)}, showing a periodicity of 0.5 GHz and 0.23 GHz, respectively. The expected periodicity due to the external cavity is given by $c/2L_\text{ext}$, which gives 0.4 GHz and 0.25 GHz, respectively.}
    \label{suppl_fig15b}
\end{figure}

\textbf{External cavity and sub-fringes spacing.} We now discuss in detail the effect of the external cavity formed by the SLG layer and the silica layer (representing the THz source) as presented in Suppl.~Fig.~\ref{suppl_fig15}. We consider two cases: \textit{i}) the case of a source-detector distance $L_\text{ext}$ of 36 cm (two mirrors case), and \textit{ii}) the case of a source-detector distance of 61 cm (four mirrors case). The simulation gives a fringe spacing of $\Delta f_1 \simeq$ 400 MHz in the first case, and about $\Delta f_2 \simeq$ 250 MHz. This trend is also recovered experimentally in Suppl.~Fig.~\ref{suppl_fig15a} and supported by performing a Fourier transform of the experimental data as shown in Suppl.~Fig.~\ref{suppl_fig15b}. We relate this effect to the constructive/destructive interferences of the reflected light back at the emitter plane, with a fringe condition matching $\Delta f=c/2L_\text{ext}$. It is important to note here that the magnitude effect of the fringes in the spectral response is only around $\pm$ 50\%, so that the main (sub-)THz absorption increase still comes from the internal sub-THz cavity. \\

\begin{figure}[!htp]
    \centering
    \includegraphics[width=\linewidth]{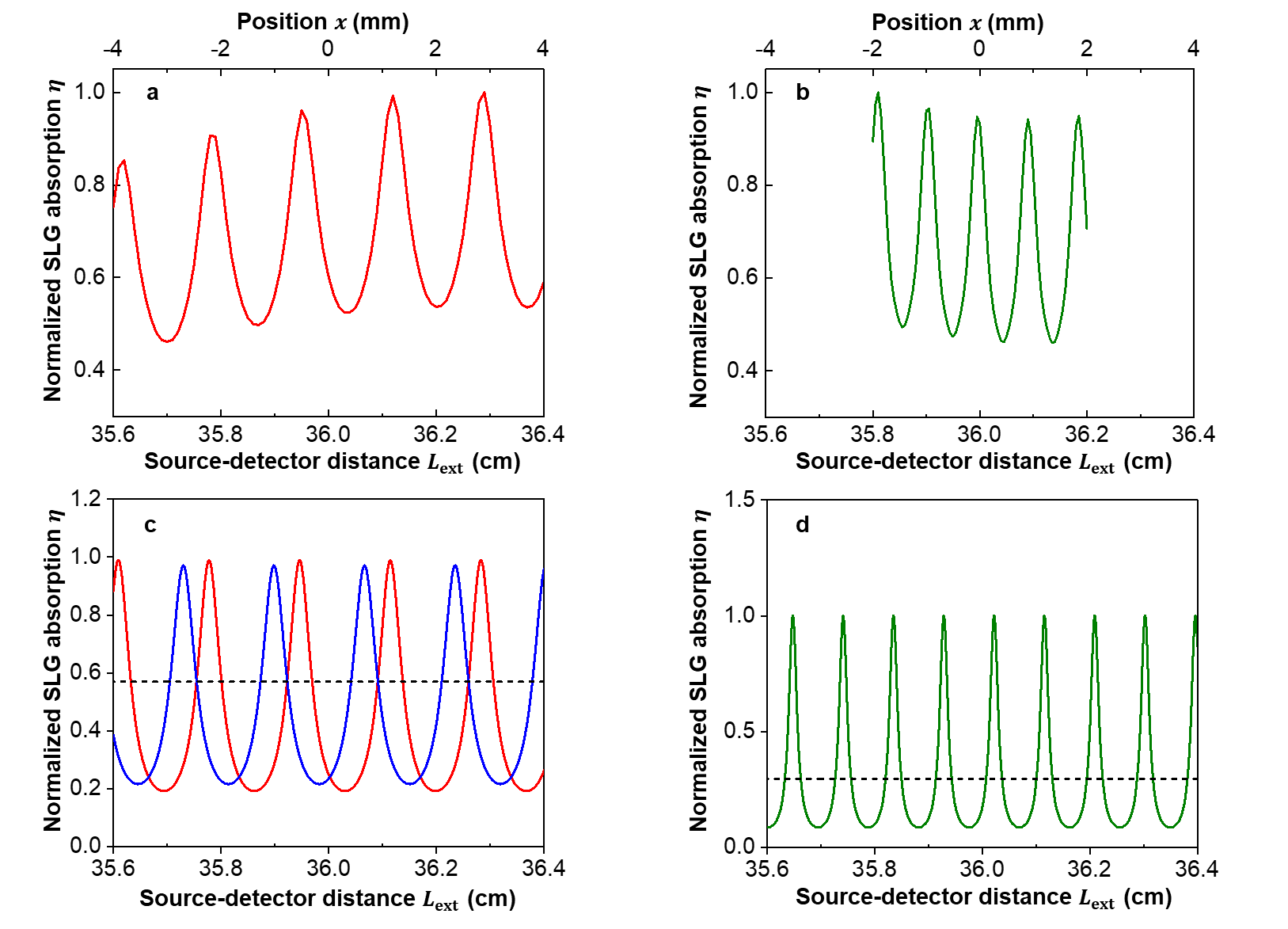}
    \caption{\textbf{Simulated detector response as a function of its position $x$}. We simulate the absorption of the SLG by moving the position of the detector along the focal plane. First, we simulate for two different frequencies: \textbf{(a)} 89 GHz and \textbf{(b)} 160 GHz. Second, we simulate \textbf{(c)} the presence of the interference pattern with a Si wafer of 500 $\mu$m between the THz emitter and the coherent graphene receiver (blue curve), without the Si wafer (red line), and the absorption of the SLG without any cavity (black dashed curve) at 89 GHz. Finally, we observe the contribution of the internal and external cavity by simulating \textbf{(d)} SLG with the internal and external cavity (green line) and the SLG without any cavity (black dashed line) at 160 GHz.}
    \label{suppl_fig24}
\end{figure}

\textbf{Response of the device \textit{vs.} position along the THz propagation path.} We present the simulated response of the device vs position $x$ for two different frequencies, 89 GHz (Suppl.~Fig.~\ref{suppl_fig24}\reb{a}) and 160 GHz (Suppl.~Fig.~\ref{suppl_fig24}\reb{b}). We observe that the oscillation period $\Delta x$ (1.69 mm for 89 GHz and 0.94 mm for 160 GHz) matches closely the one obtained experimentally. 
The effect of the internal and external cavity is then demonstrated at 160 GHz in Suppl.~Fig.~\ref{suppl_fig24}\reb{d} where we see a large increase of the SLG absorption compared to the absorption of the SLG without any cavity. Finally, we report in Suppl.~Fig.~\ref{suppl_fig24}\reb{c} the effect of inserting a 500 $\mu$m Si wafer between the source and the detector at 89 GHz, shifting the oscillation by $\Delta x$ corresponding to $d/(n_s-1)$. \\

\clearpage 
\section{Photocurrent properties under excitation with 1550 nm light}
\label{S_telecom}

\begin{figure}[!htp]
    \centering
    \includegraphics[width=0.7\linewidth]{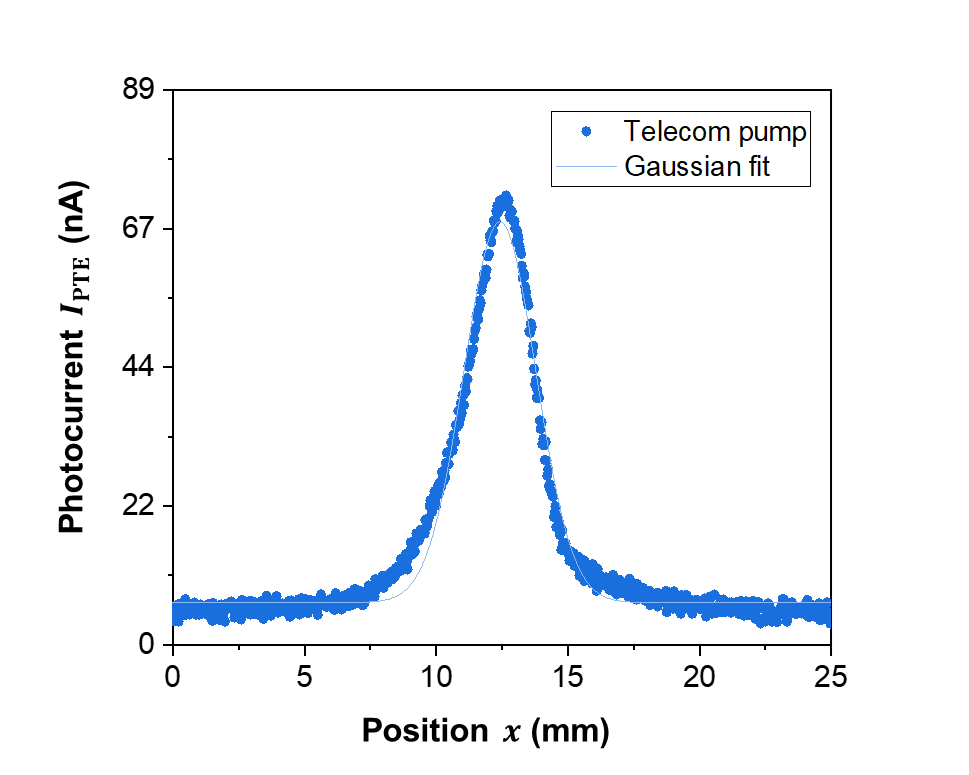}
    \caption{\textbf{Profile of the telecom photocurrent vs focal position.} Photocurrent generated from telecom pump as a function of the position across the propagation direction $x$. }
    \label{suppl_fig7}
\end{figure}

\textbf{Photocurrent \textit{vs.} position for 1550 nm light.} We report in Suppl.~Fig.~\ref{suppl_fig7} the photocurrent as a function of position along the propagation direction $x$ for light with a near-infrared (NIR), or ``telecom'', wavelength of 1550 nm. We observe, contrary to the (sub-)THz-generated photocurrent case, no photocurrent oscillations. For this wavelength, $\lambda_0/2 \simeq 775$~nm, which is too small to be observed with our minimal step size of $1.25$ µm. 
\\

\begin{figure}[!htp]
    \centering
    \includegraphics[width=0.7\linewidth]{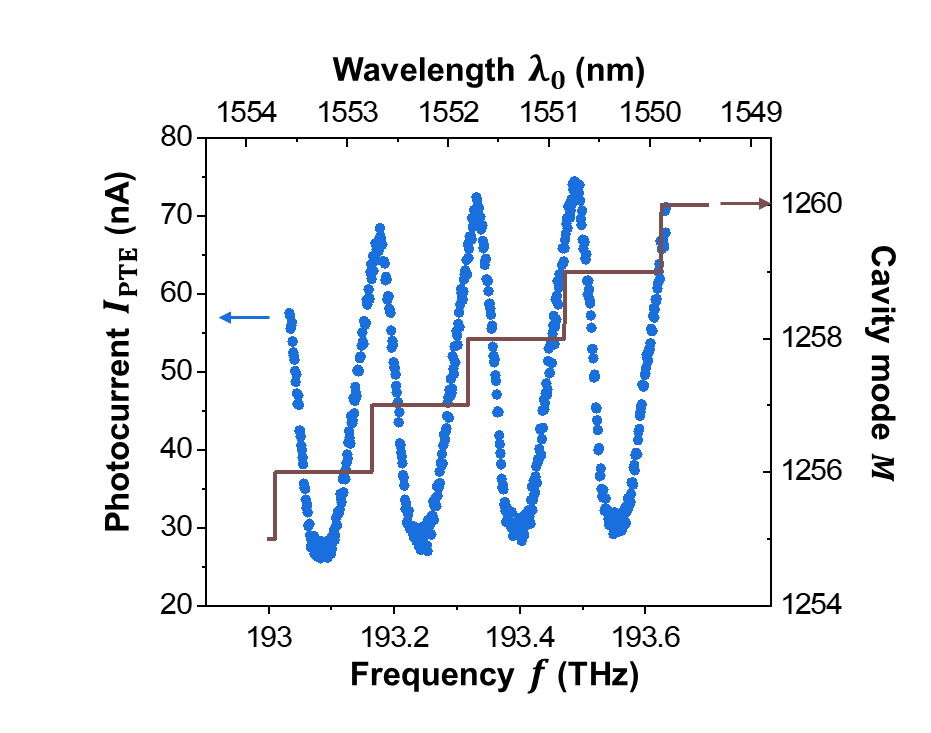}
    \caption{\textbf{Determination of the frequency response of the cavity under 1550 nm pumping.} Estimation of the number of telecom modes filling the cavity compared to the experimental measurements of the photocurrent.}
    \label{suppl_fig10}
\end{figure}

\textbf{Mode filling of the cavity with 1550 nm light.} To determine the spectral response of the cavity under telecom pumping, we assume that the sub-THz cavity of thickness $L$ possesses a refractive index of $n_{\rm NIR} \simeq 3.5$~\cite{franta_temperature-dependent_2017} at NIR/telecom frequencies, different compared to THz frequencies $n_\text{THz} \simeq 3.4$~\cite{franta_temperature-dependent_2017}, and that the cavity resonance follows the condition of $L = (2M-1) \lambda_0/4 n$. This condition is valid in both telecom and THz ranges, where $n$ is $n_{\rm NIR}$ or $n_{\rm THz}$, respectively. 
The number of modes in the cavity $M$ is expressed by the following equation: 

\begin{equation}
    M = \Bigl \lfloor \frac{2 n L}{c} f - \frac{1}{2} \Bigr \rfloor 
    \label{Seq_mode}
\end{equation}

where the operator $\floor{M}$ indicates rounding to the nearest integer that is smaller than or equal to $M$. 
When the cavity accepts/rejects a new mode, the increment matches a constructive interference. We calculate the response of the cavity under 1550 nm pumping and compare it to the experimental photocurrent in Suppl.~Fig.~\ref{suppl_fig10}. We observe a very good agreement between our experimental results with the mode filling of a sub-THz cavity. \\

\clearpage
\section{Performance of four (sub-)THz graphene detectors}
\label{S_dev}

\begin{figure}[!htp]
    \centering
    \includegraphics[width=\linewidth]{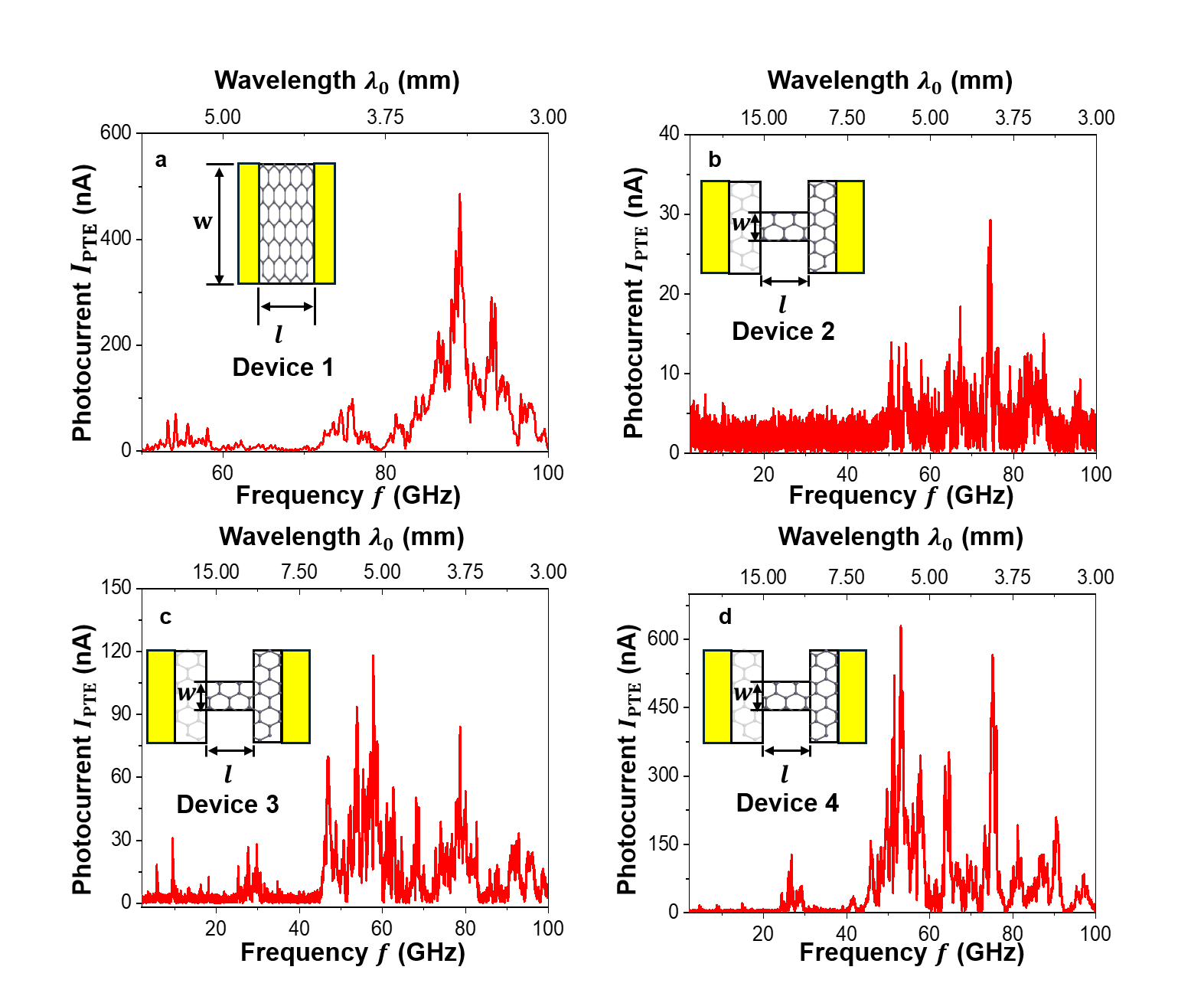}
    \caption{\textbf{Photocurrent response of four (sub-)THz detector-interferometer devices.} We measure the photocurrent $I_\text{PTE}$ \textit{vs.} frequency $f$ for the four different devices. We vary the thickness of the Si substrates to fix the resonance frequency of the cavity to (a) $\sim$90 GHz, (b), (c) and (d) $\sim$40 GHz corresponding to a thickness of the Si substrate of (a) 280 $\mu$m, (b), (c) and (d) 525 $\mu$m. We optimized the responsivity by varying the length $l$ and the width $w$ of the graphene channel, where Device 1 has $w$ = 30 $\mu$m and $l$ = 2 $\mu$m, Devices 2 and 3 have $w$ = 1 $\mu$m and $l$ = 1 $\mu$m, and Device 4 has $w$ = 3 $\mu$m and $l$ = 1 $\mu$m.}
    \label{suppl_fig12}
\end{figure}

\begin{figure}[!htp]
    \centering
    \includegraphics[width=\linewidth]{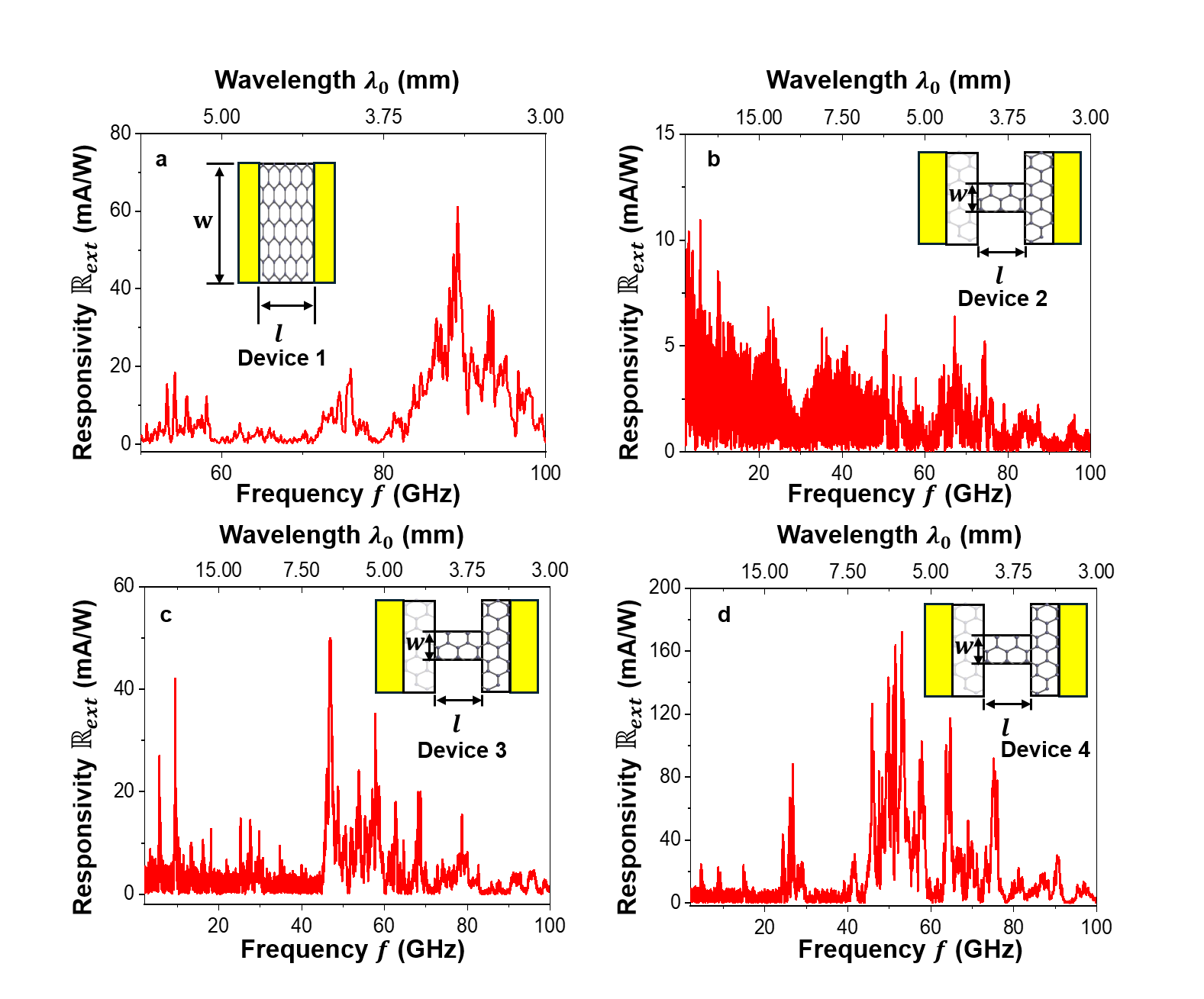}
    \caption{\textbf{External responsivity of (sub-)THz interferometers.} We determined the external responsivity $\mathbb{R}_{\rm ext}$ of the four different devices and used graphene channel resistances of (a) 500 $\Omega$, (c) 1031 $\Omega$ and (d) 830 $\Omega$ to obtain the NEP, achieving the lowest NEP of 26 pA/$\sqrt{\rm Hz}$  for Device 4 in panel d.}
    \label{suppl_fig12}
\end{figure}

\clearpage
\section{Comparison of external performances of graphene-based free photodetectors}
\label{S_table}

We present in Suppl.~Table~\ref{extended_1} a comparison of the performance of different graphene-based THz detectors. In order to compare the performance, we calculate the external responsivity of the detectors using the information provided in each reference. Since the responsivity is calculated by different approaches across the literature~\cite{javadi_sensitivity_2020}, such as the geometry of active area of the antenna, or diffraction-limited spot size, we decide to use the external responsivity as the factor of merit of those detectors, which allows for easier comparison of the performances throughout different setups and detector designs.

\begin{table*}[!htp]
    \centering
    \renewcommand{\arraystretch}{1.2} 
    \resizebox{\textwidth}{!}{%
    \begin{tabular}{|>{\centering\arraybackslash}m{3cm}|>{\centering\arraybackslash}m{3.5cm}|>{\centering\arraybackslash}m{3cm}|>{\centering\arraybackslash}m{3cm}|>{\centering\arraybackslash}m{3cm}|>
    {\centering\arraybackslash}m{3cm}|>{\centering\arraybackslash}m{2.5cm}|}
    
    \hline
    \textbf{Device design} & \textbf{Frequency} & \textbf{Responsivity} & \textbf{NEP} & \textbf{Reference} \\ \hline

    Square-spiral antenna without cavity 
    & 80–120, 140, 300 GHz 
    & 1.4 mA/W** $\quad$(2.8 mV/W*) 
    & 35 mW~$\rm{Hz}^{-1/2}$ 
    & Guo~\textit{et al.}~\cite{guo_graphene-based_2018} \\ \hline

    Log-periodic antenna 
    & 120 GHz 
    & 0.84 mA/W** $\quad$(840 mV/W*) 
    & 33.3 nW~$\rm{Hz}^{-1/2}$* 
    & Liu~\textit{et al.}~\cite{Liu2018} \\ \hline

    Dipolar antenna without cavity 
    & 1.8–4.25 THz 
    & 0.42 mA/W* $\quad$(1.76 mV/W**) 
    & 4.8 nW~$\rm{Hz}^{-1/2}$*
    & Castilla~\textit{et al.}~\cite{castilla_fast_2019}\\ \hline
    
    Bow-tie antenna with cavity 
    & 2.86 THz 
    &$\approx$ 0.1 mA/W** ~~~($\approx$ 2 V/W*) 
    & 5.4 nW~$\rm{Hz}^{-1/2}$*
    & Viti~\textit{et al.}~\cite{Vitiello2025Salesburry}\\ \hline

    Integrated (sub-)THz cavity and dipolar antenna (Device 1)
    & 89 GHz 
    & 73 mA/W $\quad \quad \quad$(36 V/W) 
    & 79 pW~$\rm{Hz}^{-1/2}$
    & This work \\ \hline
    
    Integrated (sub-)THz cavity and bow-tie antenna (Device 4)
    & 53 GHz 
    & 172 mA/W $\quad \quad \quad$(138 V/W) 
    & 26 pW~$\rm{Hz}^{-1/2}$
    & This work \\ \hline
    \end{tabular}}
    \caption{\textbf{Comparison of the~\textit{external} performance of graphene-based, bias-free photodetectors.}
    *Maximum external responsivities calculated considering the experimental focus area using the normalization factor provided in the referenced papers. **These values were converted using the values reported in the referenced papers, such that they correspond to values without normalization.}
    \label{extended_1}
\end{table*}

\clearpage
\section{Time-resolved photocurrent measurements}
\label{S_TRPC}

\textbf{Reflectance and photocurrent map.} We present in Suppl.~Fig.~\ref{suppl_fig12} the reflectance and photocurrent maps of the device in the region of the antenna/graphene region obtained using only the pump beam. We observe the geometry of the device in Suppl.~Fig.~\ref{suppl_fig12}\reb{a} and focus on the region of interest. In Suppl.~Fig.~\ref{suppl_fig12}\reb{b}, we show the photocurrent measured by moving the device with the $xyz$ stage. We observe three photocurrent spots. By comparing their positions to the reflectance map, we realize that two of them originate from the photoactive region, while the last one is located outside the graphene channel. This is evidence of photocurrent that is induced by incident light that is absorbed in graphene via a reflection of the metallic back mirror. 

\begin{figure}[!htp]
    \centering
    \includegraphics[width=\linewidth]{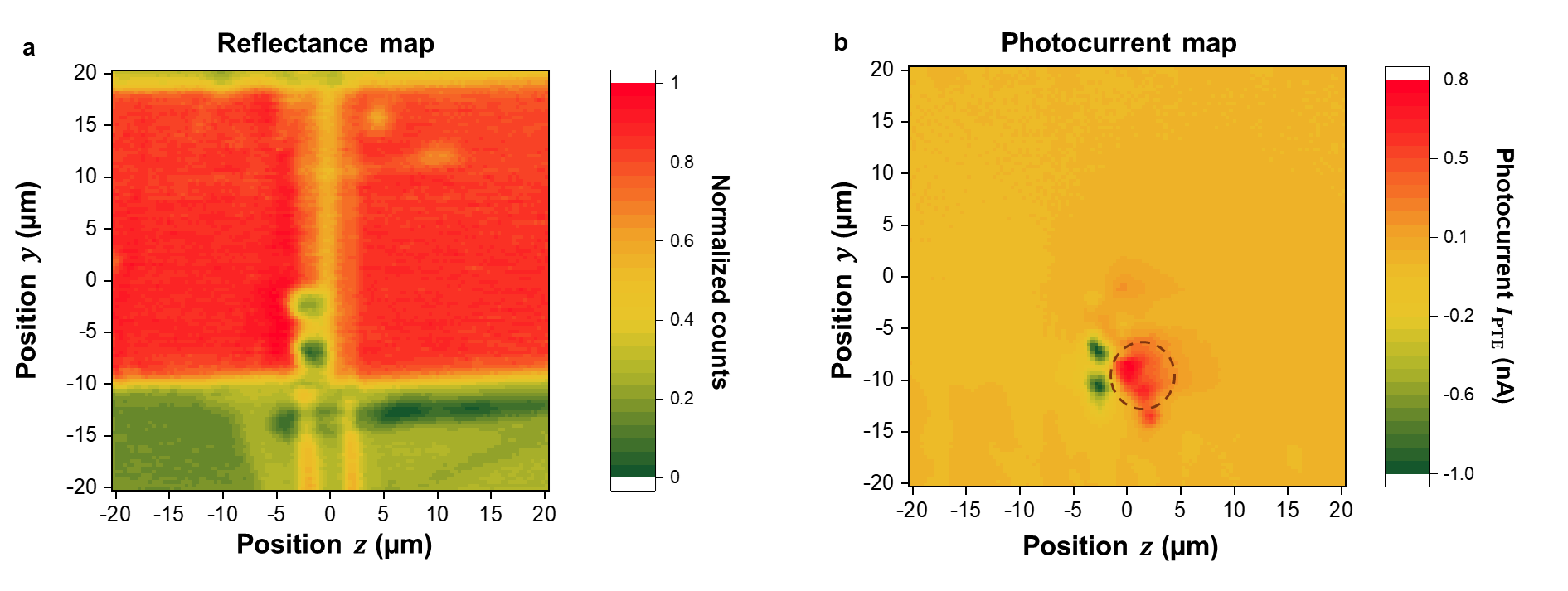}
    \caption{\textbf{Time-resolved photocurrent map.} \textbf{a,} Reflection map and \textbf{b,} photocurrent map in the region of the antenna/graphene channel. The dashed black circle indicates the spot where the photocurrent dynamics were measured.}
    \label{suppl_fig12}
\end{figure}

\begin{figure}[!htp]
    \centering
    \includegraphics[width=0.7\linewidth]{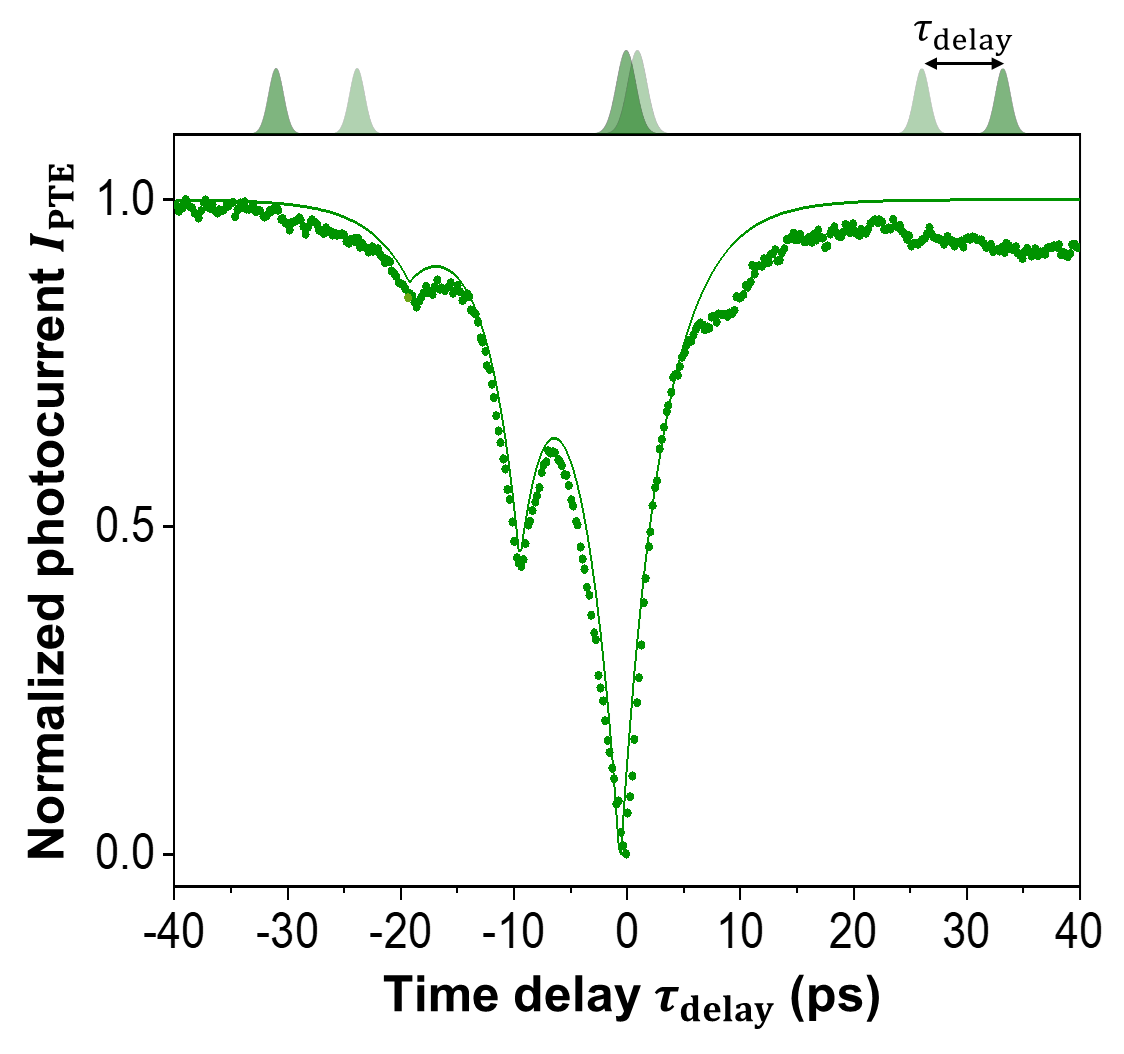}
    \caption{\textbf{Time-resolved photocurrent measurements.} PTE photocurrent measured using time-resolved photocurrent technique (green points) and the sum of exponential decay function (solid line). We observe a main dip at zero time delay followed by two  secondary periodic dips related to reflections inside the (sub-)THz cavity.   
    }
    \label{suppl_fig12a}
\end{figure}

\textbf{Determination of the cooling time from the TRPC measurements.} In order to determine the cooling time of the hot-electrons in the graphene THz detector, we perform time-resolved photocurrent measurements with an incident wavelength of 1030 nm. For details on the technique, see \meths~and Ref.~\cite{tielrooij_generation_2015}. To measure the cooling time, we use a time-resolved photocurrent technique explained in \meths ~section. We used a power of around 600 $\mu$W for both lasers and record the PTE photocurrent as shown in Suppl.~Fig.~\ref{suppl_fig12a}. We observe a pronounced main dip at zero time delay, and secondary periodic dips that are likely related to reflections inside the (sub-)THz cavity. To extract the cooling time from the time-resolved photocurrent, we describe the photocurrent using a sum of exponential decay functions of the form $ I_\text{PTE} \propto (T_e - T_l) \exp \left( -\frac{\tau_\text{delay}} {\tau_\text{cool}}\right)$~\cite{tielrooij_hot-carrier_2015,tielrooij_generation_2015}, where $T_e$, $T_l$, $\tau_\text{delay}$, $\tau_\text{cool}$ are the electronic temperature, lattice temperature, the delay time, and the cooling time, respectively. Each dip has two exponentially decaying functions -- to positive and to negative time delays. We obtain a cooling time of 3.4 ps for the main dip around time zero.

\clearpage

\section{Phase sensitivity}
\label{PhaseSensitivity}

\begin{figure}[!htp]
    \centering
    \includegraphics[width=\linewidth]{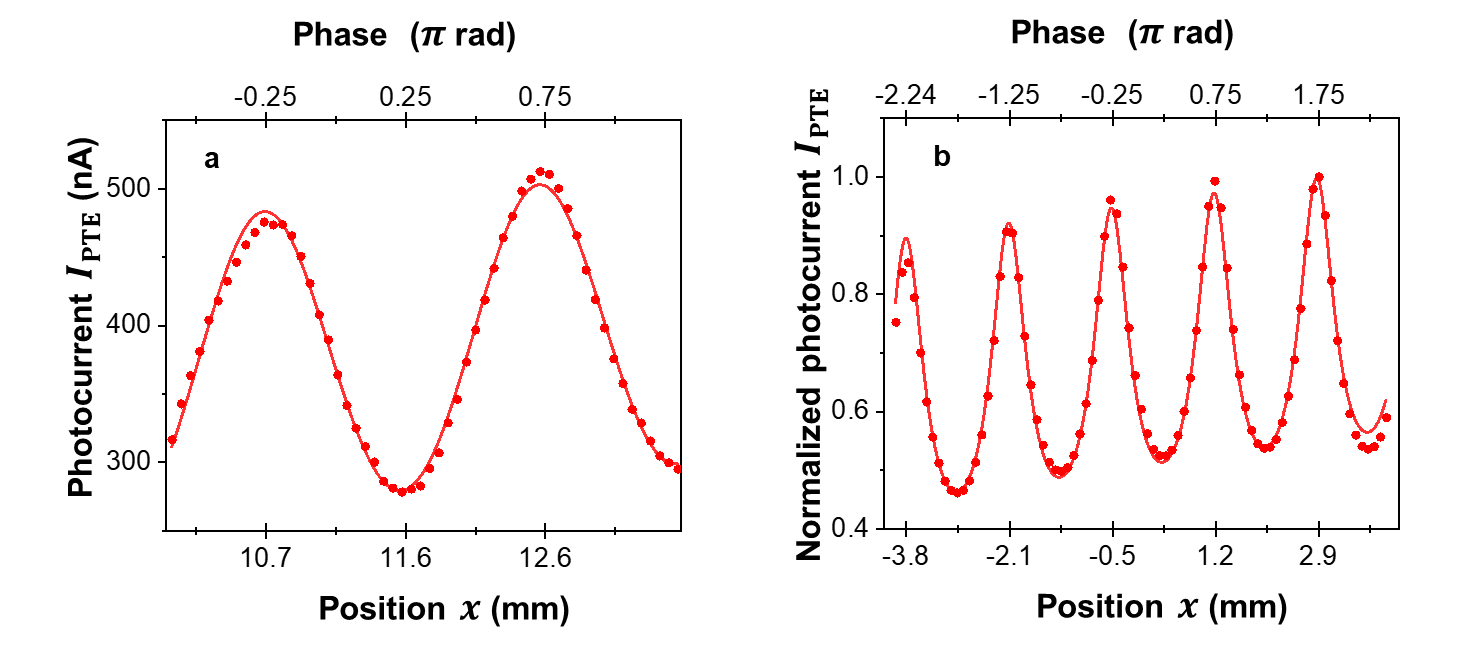}
    \caption{\textbf{Airy function of oscillation photocurrent pattern.} We describe the photocurrent oscillation pattern (red points) using an Airy function (solid lines) for the (a) experimental measurements and (b) simulation. 
    }
    \label{suppl_fig28}
\end{figure}

\begin{figure}[!htp]
    \centering
    \includegraphics[width=0.7\linewidth]{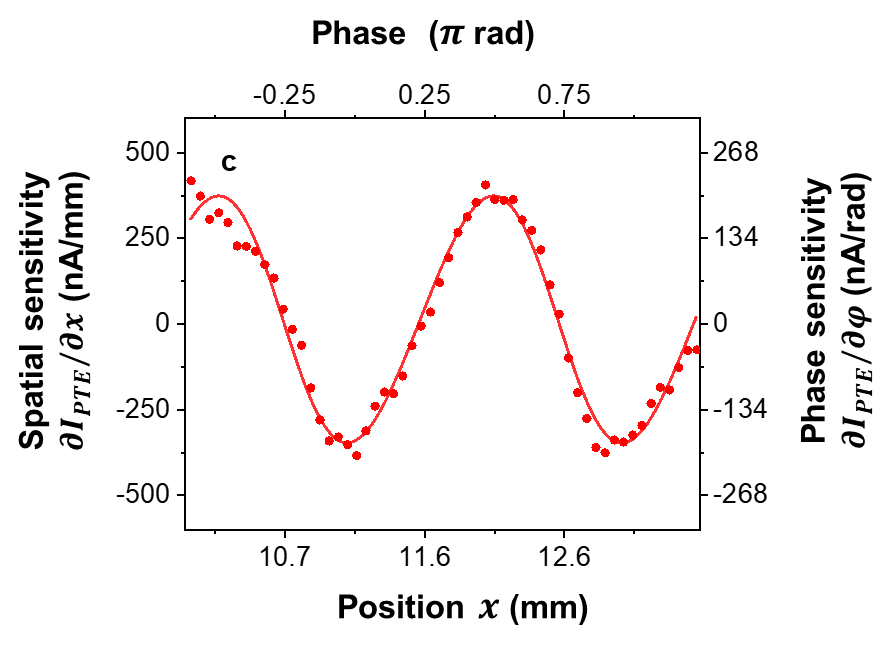}
    \caption{\textbf{Spatial and phase sensitivity.} We extract spatial sensitivity ($\partial I_\text{PTE}/\partial x$) and phase sensitivity ($\partial I_\text{PTE}/\partial \varphi$) of our sub-THz interferometer by performing the numerical derivative in Suppl. Fig. \ref{suppl_fig28}\reb{a} for experimental data (points) and fitted Airy function (solid line).
    }
    \label{suppl_fig29}
\end{figure}

\textbf{Determination of phase sensitivity} The photocurrent could be described as typical Airy function of a Fabry-Perot cavity like $I_\textbf{PTE}=I_0 \frac{1}{1+Fsin^2\varphi}+I_\text{offset}$, where $I_0$ is the maximum photocurrent, $F$ is the coefficient of the finesse, $\varphi=2\pi x/\lambda_0$ is the phase accumulation and $I_\text{offset}$ is the offset of the photocurrent due to the envelope of the THz Gaussian beam. We observed good agreement by fitting the experimental data and the simulation of the photocurrent oscillation in Suppl. Fig. \ref{suppl_fig28}\reb{a},\textbf{\reb{b}}, extracting a finesse coefficient of 0.05 using the experimental data and 2 using the simulation data. To obtain the spatial sensitivity, we applied a numerical derivative to the experimental and fitting data in Supp. Fig. \ref{suppl_fig28}\reb{a}, calculating a maximum phase sensitivity of 375 nA/mm as shown in Supp. Fig. \ref{suppl_fig29} . We transform spatial to phase sensitivity using $\frac{\partial I_\text{PTE} }{\partial \varphi}=\frac{\partial I_\text{PTE}}{\partial x} \cdot \frac{\partial x}{\partial \varphi} $ where $\frac{\partial x}{\partial \varphi} =\frac{\lambda_0}{2\pi}$. 

\label{S_airy}

\begin{figure}[!htp]
    \centering
    \includegraphics[width=0.7\linewidth]{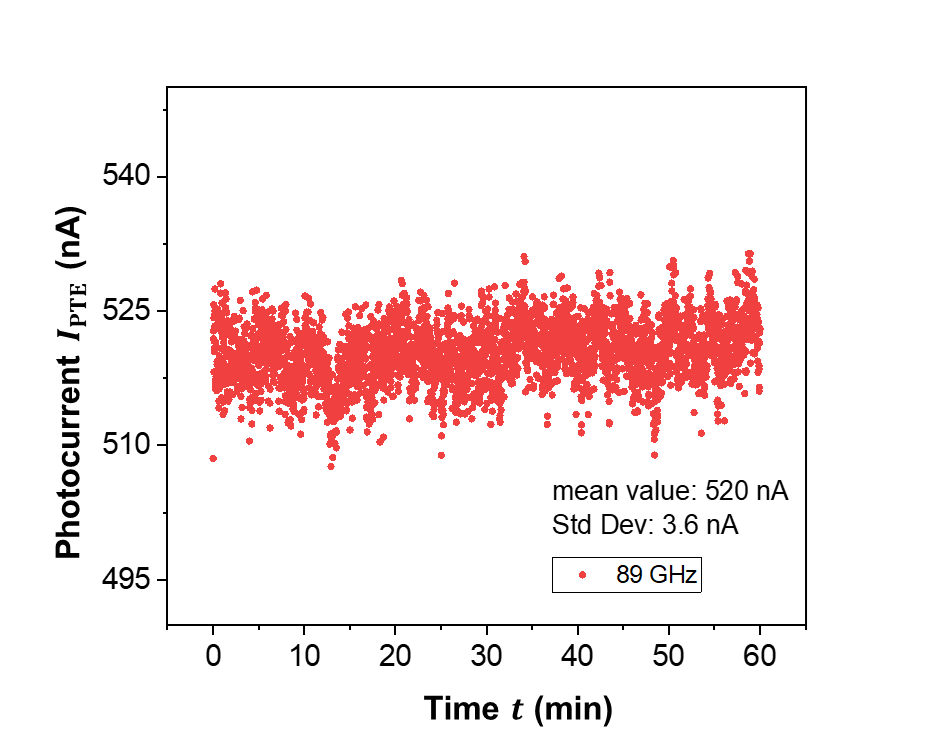}
    \caption{\textbf{Evaluation of the graphene detector noise.} Photocurrent variation as a function of time $t$ at 89 GHz using our graphene-based (sub-)THz detector-interferometer, measured during a period that is equal to the time it takes to perform a thickness measurement. The noise of the signal is 3.6 nA, extracted as the standard deviation from the root mean square value of the photocurrent.}
    \label{suppl_fig20c}
\end{figure}

\textbf{Determination of noise photocurrent} During the extraction of the thicknesses, potential uncertainties entering into error bars could intervene. The first main effect is the phase drift of the THz system, that relies on two photomixed DFB lines to produce a THz beating note. A frequency drift of each DFB due to thermal fluctuations would here result in a source of error. We next measure at 89 GHz for one hour, as shown in  Suppl.~Fig.~\ref{suppl_fig20c} and calculate the noise photocurrent related to the phase drift of our THz source by extracting the standard deviation of the photocurrent along this period of time.\\

\begin{figure}[!htp]
    \centering
    \includegraphics[width=1\linewidth]{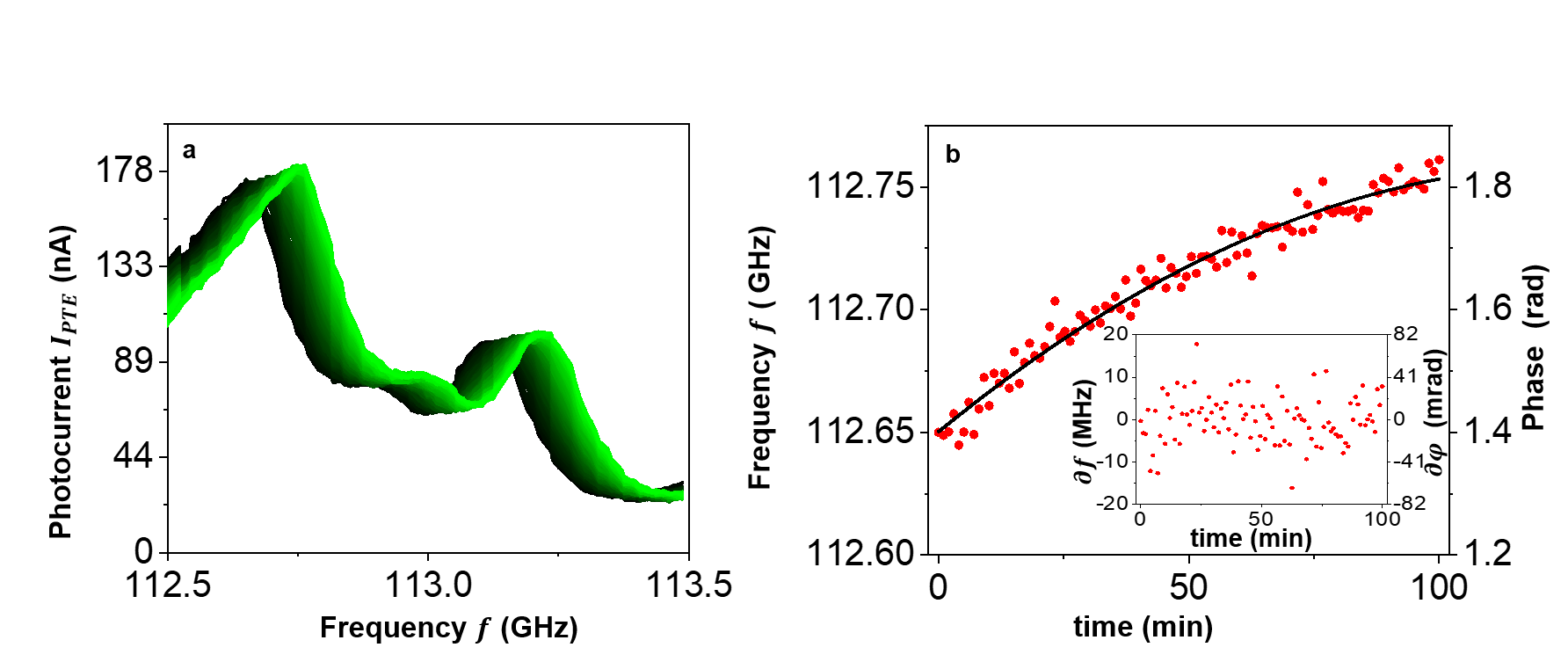}
    \caption{\textbf{Determination of frequency and phase drift of the THz source.} (a) Temporal evolution of the frequency response of our (sub-)THz detector. (b) Frequency drift extracted by locking the frequency at the maximum $I_\text{PTE}$ (red points) in (a) and the phase drift tendency (black solid line) by fitting a polynomial fit. The inset shows the frequency fluctuation on top of the drift, obtained by subtracting the long-term frequency drift. The standard deviation of these fluctuations is around 5 MHz. }
    \label{suppl_fig30}
\end{figure}

\textbf{Determination of phase drift} We determine the phase of the THz wave by measuring a small frequency window with our (sub-)THz interferometer with an interval period of one minute as shown in Suppl.~Fig.~\ref{suppl_fig30}\reb{a}. We then extracted the resonance frequency of each scan by looking at the maximum photocurrent value. We observe in Suppl.~Fig.~\ref{suppl_fig30}\reb{b} a frequency drift, most likely due to the variation of temperature of the DBF lasers. This drift can lead to a frequency change up to 100 MHz for a measurement of one hour. To extract the frequency fluctuations of our THz source, we subtract the general tendency of the frequency drift by fitting with a polynomial function. The remaining fluctuations have a standard deviation of $\sim$5 MHz. To extract the relation between the frequency drift and the phase drift, we use that phase and frequency of the external cavity are related via $\varphi=\pi L f/c $, where $L$ is the length of the cavity between the THz source and the detector (36 cm), and $f$ the frequency of the THz source. We thus obtain that the phase noise is  $\delta \varphi=\pi L \delta f/c $, where $\delta f$ is the frequency fluctuation. This gives a phase noise of 0.018 rad, which is very similar to experimentally obtained phase accuracy.

\clearpage

\section{Additional measurements for the thickness determination}
\label{S_imaging}

\begin{figure}[!htp]
    \centering
    \includegraphics[width=\linewidth]{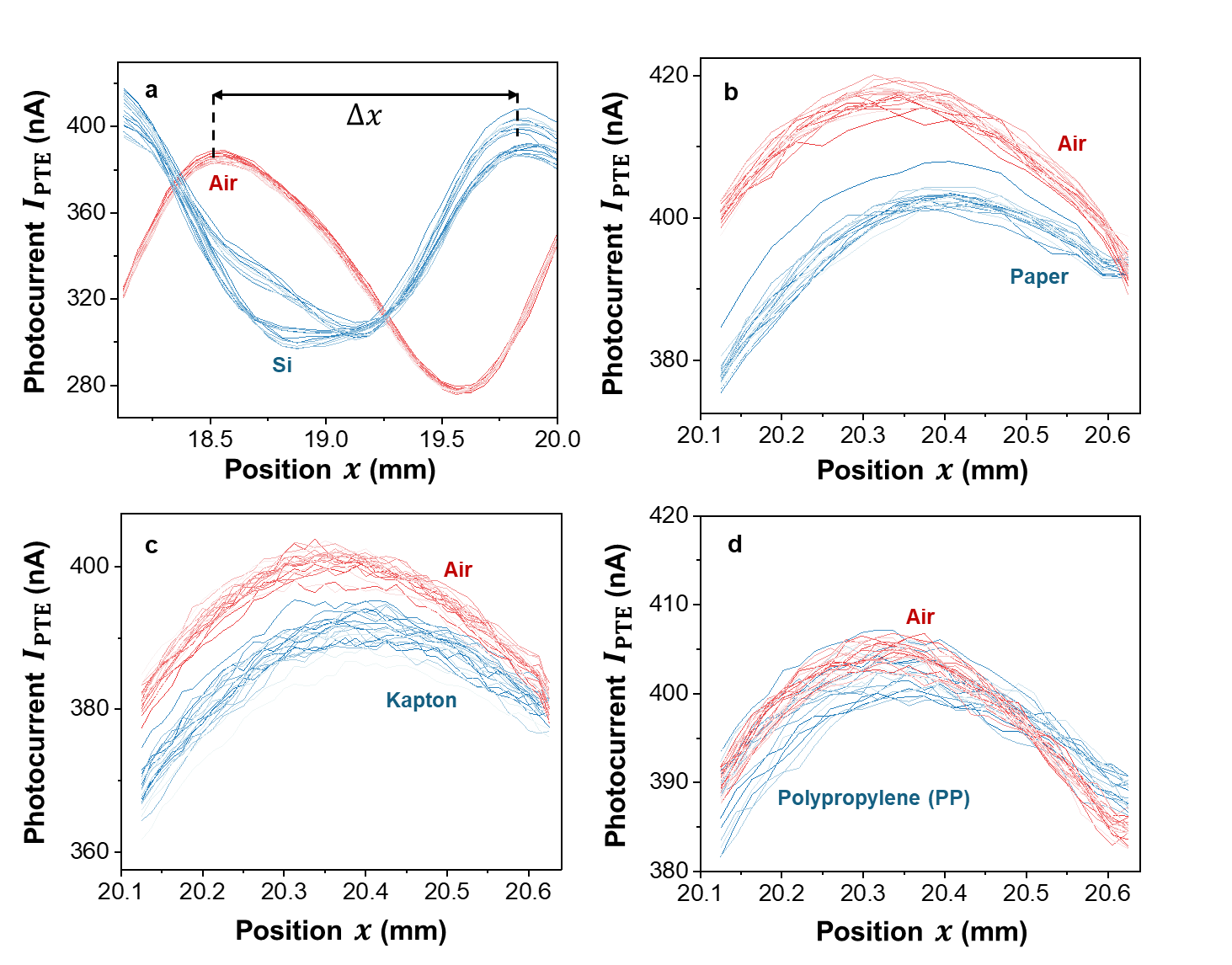}
    \caption{\textbf{Cycles of measurements for the thickness estimation.} Complete measurements for photocurrent $I_\text{PTE}$ vs position $x$ of the detector along the THz propagation direction, without (red solid lines) and with the sample (blue solid lines) in front of the detector, for \textbf{(a)} Si, \textbf{(b)} Paper, \textbf{(c)} Kapton, and \textbf{(d)} Polypropylene (PP).}
    \label{suppl_fig20}
\end{figure}

\textbf{Thickness extraction method.} To extract the thickness of different materials, we choose four different materials with different refractive indices ($n_s$): silicon, paper, scotch tape (polypropylene, PP) and Kapton tape (polyimide). We first measure the photocurrent $I_\text{PTE}$ along the THz propagation axis $x$ to observe around three oscillations without the material in front of the detector (air case) and with the material in front of it. Then, we estimate the range of the shift $\Delta x$ (with respect to the $xyz$ stage position) to know the limits where we need to move the stage to observe two peaks (with and without the material). Because our detector is very sensitive to phase changes, the intrinsic phase drift of our THz emitter creates a mismatch in the measurement. To solve this problem, we decrease the scanning window using our previous estimation of how long we need to move the stage to observe the two peaks, leading to a faster measurement in terms of time, thus minimizing the effect of potential phase drift. We measure the photocurrent without the material in front of the detector and then with the material in front of the detector. To increase the accuracy of our measurements, we perform twenty cycles to enhance our statistical analysis as shown in Suppl.~Fig.~\ref{suppl_fig20}. \\

To calculate the thickness, we apply two different methods to the same dataset: \textit{i}) the first method calculates the shift $\Delta x$ (as discussed in the main text) by searching the maximum value of the consecutive scans and calculating the thickness $d$ of each cycle (with and without the material) via $d = \Delta x /(n_s-1)$. The second method \textit{ii}) first performs a smoothing of each measurement in the dataset with a window of eleven points, and then we estimate the thickness in the same way as discussed for the first method. For both methods, the twenty estimated thicknesses are averaged and the standard deviation (error bars) is calculated. \\

\begin{figure*}[!htp]
    \centering
    \includegraphics[width=0.9\linewidth]{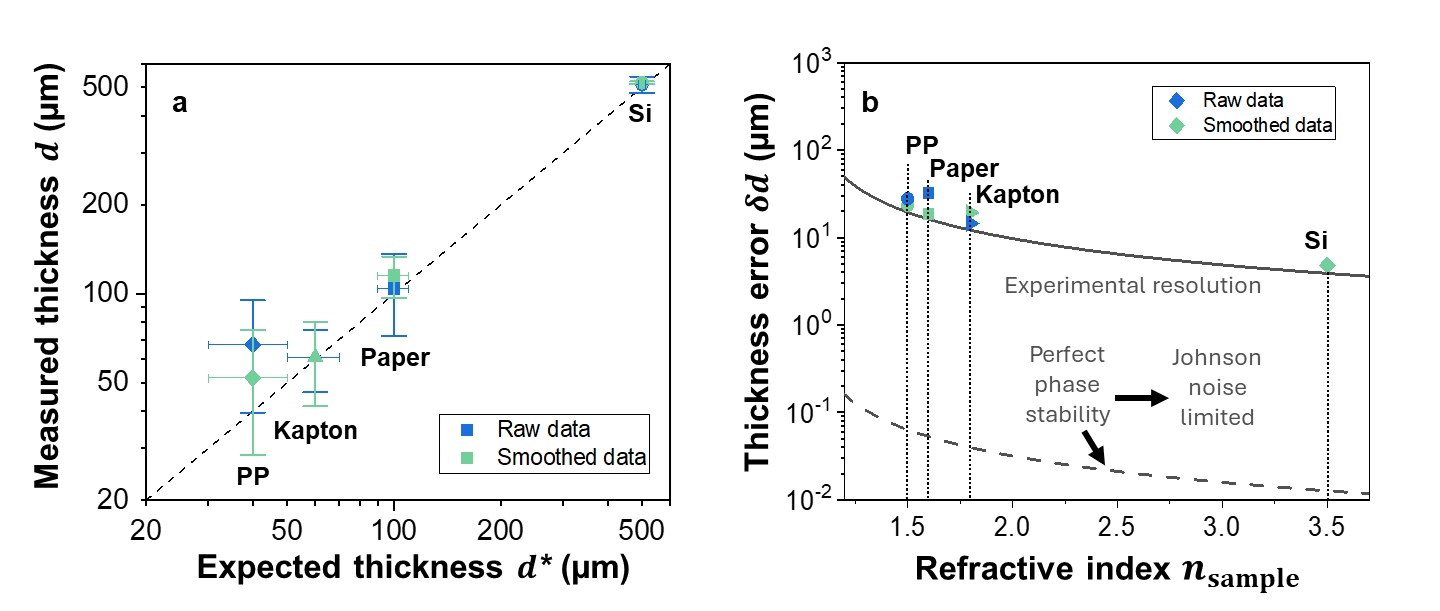}
    \caption{\textbf{Comparison of thickness estimation using two different methods.} \textbf{a,} Extracted thickness $d$ \textit{vs.} expected thickness $d^*$ using the raw data (open blue symbols) and smoothed data (solid green symbols).  \textbf{b,} Obtained thickness accuracy $\delta d$ as a function of refractive index for the different materials, using the raw data (open blue symbols) and smoothed data (solid green symbols).}
    \label{suppl_fig20b}
\end{figure*}

We present in Suppl.~Fig.~\ref{suppl_fig20b} the extracted thickness and standard deviation from the two methods discussed above. The blue points correspond to the raw data method (method (\textit{i})) and the green points correspond to the smoothed data method (method (\textit{ii})). The method presented in the main text is the smoothed data method. \\

\clearpage
\section{Experimental methods}
\label{S_methods}
\begin{figure}[!htp]
    \centering
    \includegraphics[width=0.7\linewidth]{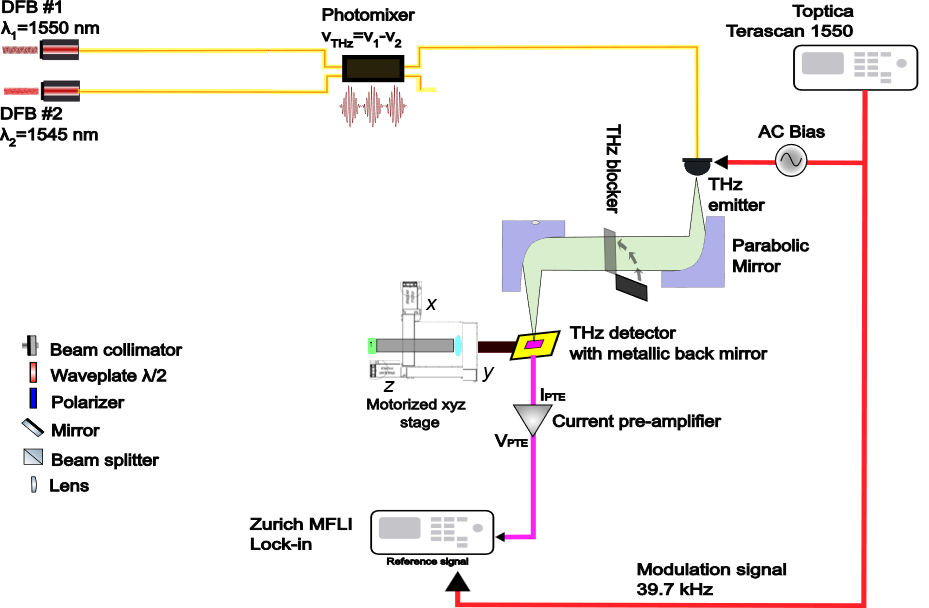}
    \caption{\textbf{Terahertz setup.} Continuous wave (CW) THz radiation is emitted by pumping an InGaAs emitter with a photomixed telecom beam, while being biased by a AC voltage, serving as a frequency reference for lock-in measurement. The CW THz beam is focused using two parabolic mirrors, and the detector is placed on an $xyz$ motorized stage to optimize its position. Finally, the photocurrent is measured using a lock-in technique demodulated at the frequency of the AC bias voltage applied to the THz emitter.}
    \label{suppl_fig3}
\end{figure}

\begin{figure}[!htp]
    \centering
    \includegraphics[width=0.9\linewidth]{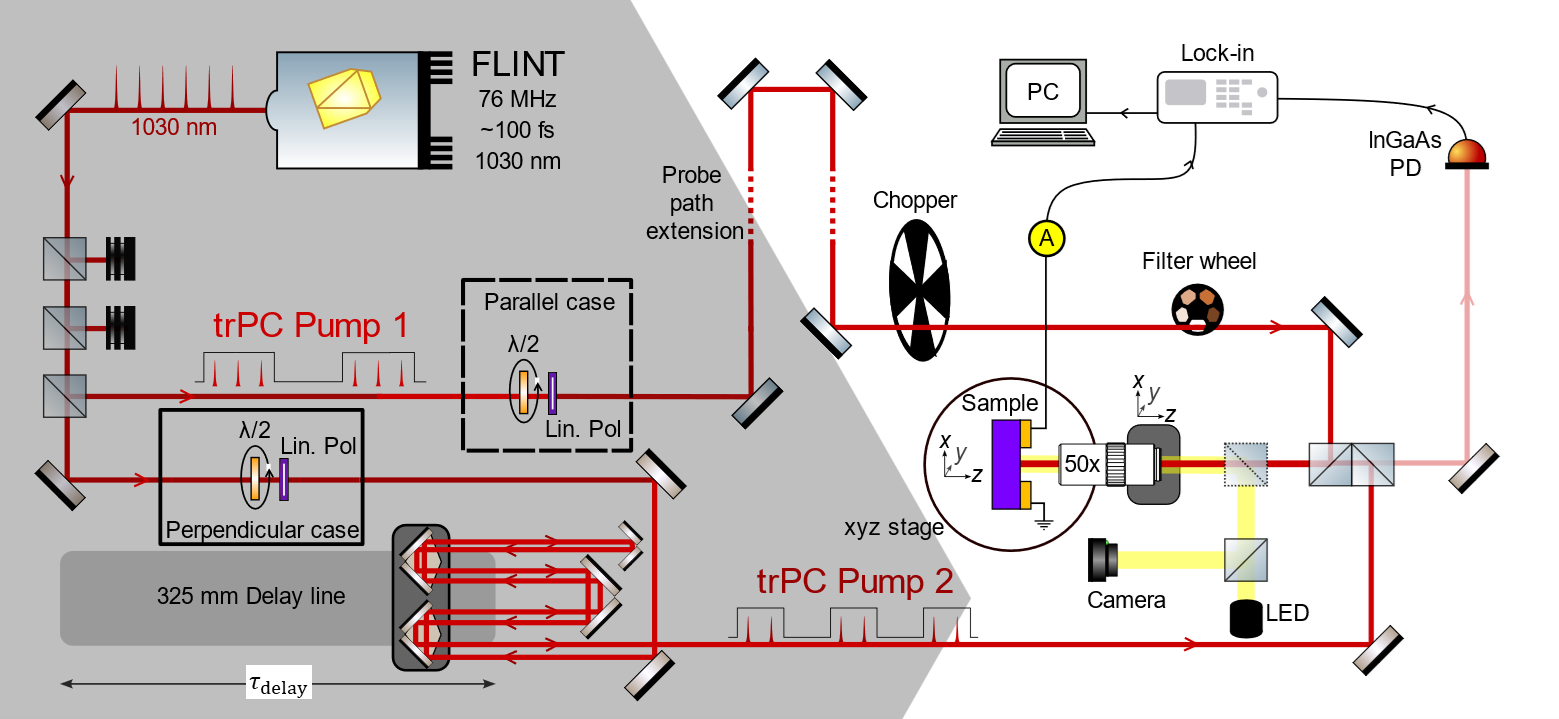}
    \caption{\textbf{Time-resolved photocurrent setup.} The time-resolved photocurrent (TRPC) setup uses a femtosecond laser with a 76 MHz repetition rate and a wavelength of 1030 nm, as explained in the \meths. We divide the beam into two paths using a beam splitter, one of the paths goes directly to an objective and the second through a 325 mm delay line to reach the objective, which focuses both beams on the sample.}
    \label{suppl_fig4}
\end{figure}

\begin{figure}[!htp]
    \centering
    \includegraphics[width=\linewidth]{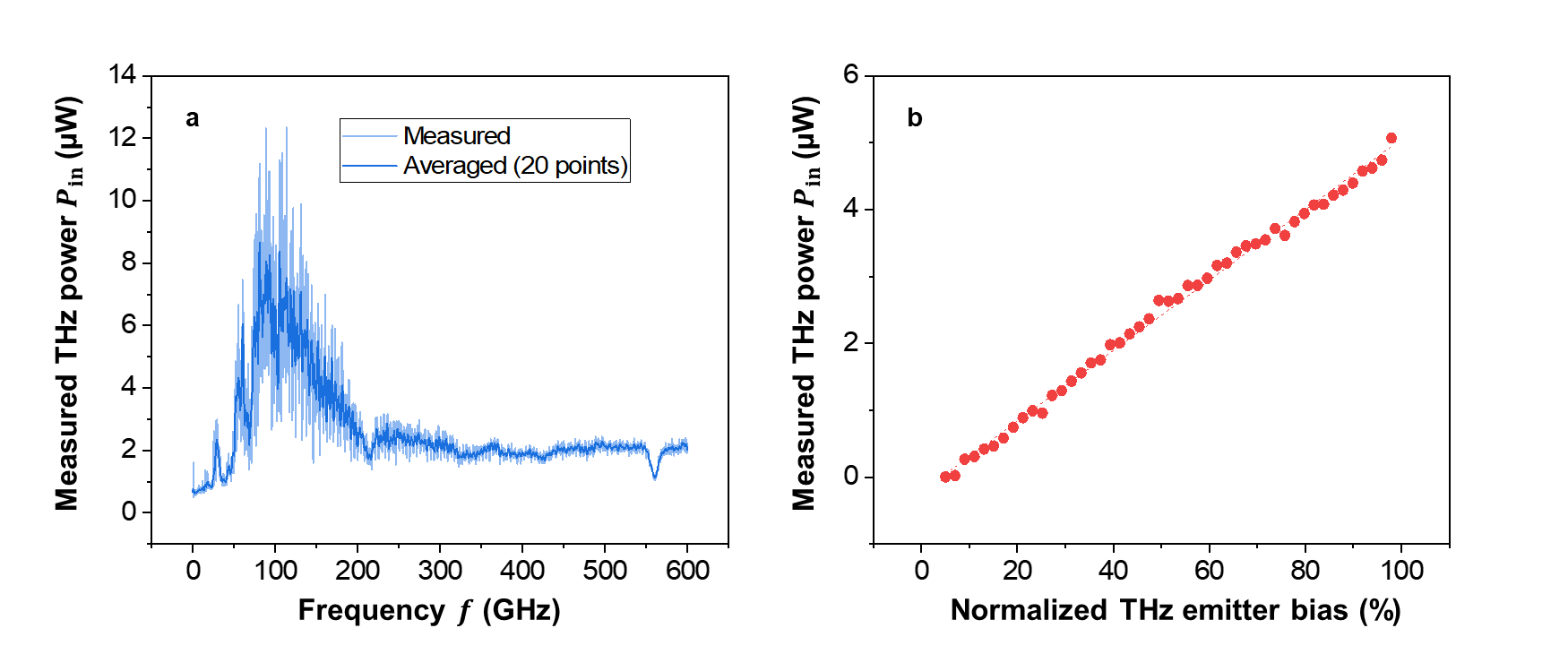}
    \caption{\textbf{Calibration of the THz source power.} \textbf{a,} Evolution of the THz power as a function of the frequency. \textbf{b,} Measured THz power at 89 GHz as a function of the emitter bias. The THz power has been measured at the THz focal point (device location) with a pyroelectric sensor (Ophir RM9-THz). A polystyrene foam has been placed in the collimated beam path to remove any residual telecom pump.}
    \label{suppl_fig5}
\end{figure}

\begin{figure}[!htp]
    \centering
    \includegraphics[width=\linewidth]{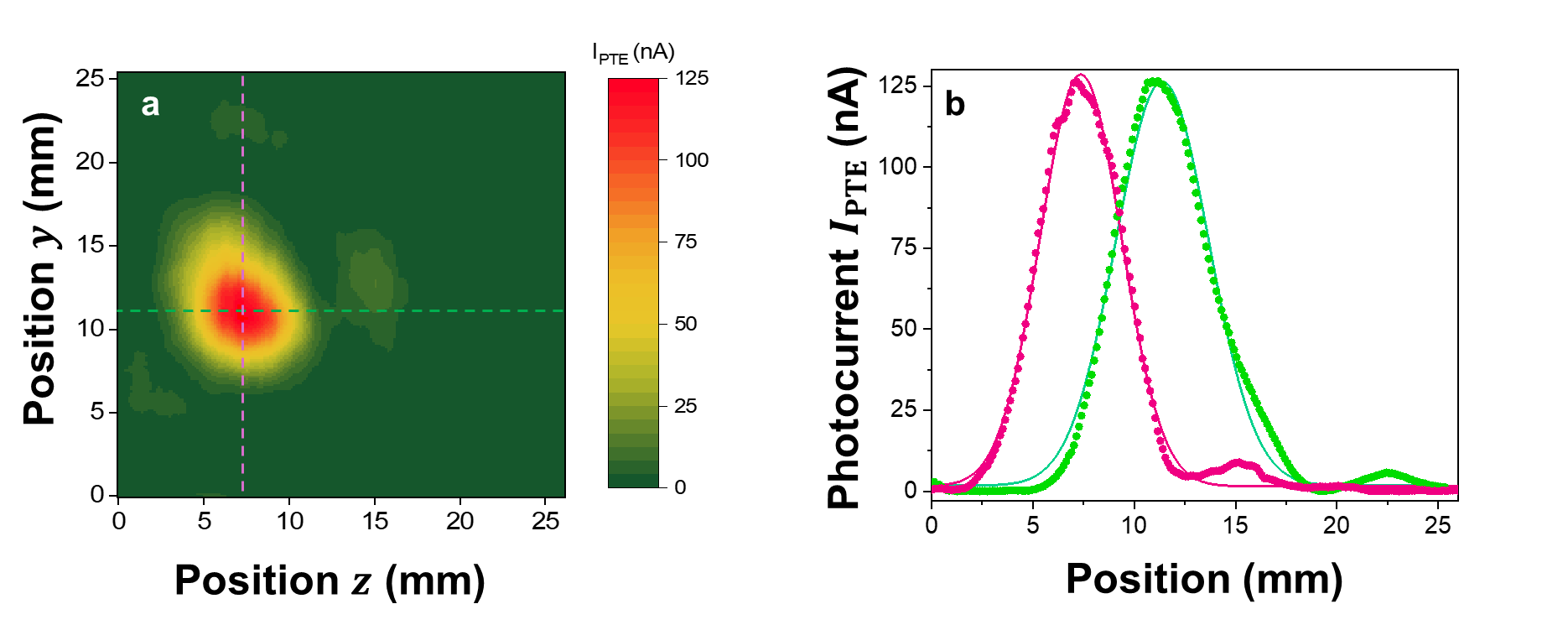}
    \caption{\textbf{Determination of the experimental focus area.} \textbf{a,} Photocurrent profile when scanning the detector position in the plane perpendicular to the travel direction of light. This is taken at 80 GHz at the best focal distance after the second parabolic mirror, and gives the experimental focus spot size. 
    \textbf{b,} Determination of the focus spot size by describing the photocurrent extracted along the vertical ($y$) and horizontal ($z$) lines (purple and green data points, respectively) with Gaussian functions (solid lines). We obtain a beam radius $w_{0,\rm y}$ = 4.58 mm and $w_{0,\rm z}$ = 4.01 mm, corresponding to a full-width at half-maximum of $\text{FWHM}=\sqrt{2 \ln{2}} ~w_0$. We calculate the focus area as $A_{\text{focus}}=\pi (w_{0,\rm z}w_{0,\rm y})$ giving $A_{\text{focus}} = 57.8 $ mm$^2$.}
    \label{extended_2}
\end{figure}

We show in Suppl.~Fig.~\ref{suppl_fig3} and Suppl.~Fig.~\ref{suppl_fig4} the schematic of the THz and time-resolved photocurrent setups (see~\meths). The THz power of the emitter as a function of the frequency is measured at the sample position with a calibrated pyroelectric THz detector (Suppl.~Fig.~\ref{suppl_fig5}\reb{a}) and the power of the THz beam as a function of the emitter bias is shown in Suppl.~Fig.~\ref{suppl_fig5}\reb{b}. We show our THz beam by measuring the photocurrent along the $yz$ plane in Suppl.~Fig.~\ref{extended_2}\reb{a}. We also determine the size of our beam by measuring the Gaussian profile of our photocurrent along the axes y and z, as shown in Suppl.~Fig.~\ref{extended_2}\reb{b}.

\clearpage
\renewcommand\refname{Supplementary References}
\bibliography{sn-bibliography}